\documentclass[aps,  prd, twocolumn,preprintnumbers,amsmath,amssymb,superscriptaddress, longbibliography]{revtex4-1}

\usepackage{import}
\usepackage{preamble}

\begin{document}

\title{Scalable repeater architectures for multi-party states}

\author{V. V. Kuzmin}
\email[E-mail me at: ]{viacheslav.kuzmin@uibk.ac.at}
\affiliation{Center for Quantum Physics, Faculty of Mathematics, Computer Science and Physics, University of Innsbruck, A-6020, Innsbruck, Austria}
\affiliation{Institute for Quantum Optics and Quantum Information of the Austrian Academy of Sciences, A-6020 Innsbruck, Austria}

\author{D. V. Vasilyev}
\affiliation{Center for Quantum Physics, Faculty of Mathematics, Computer Science and Physics, University of Innsbruck, A-6020, Innsbruck, Austria}
\affiliation{Institute for Quantum Optics and Quantum Information of the Austrian Academy of Sciences, A-6020 Innsbruck, Austria}

\author{N. Sangouard}
\affiliation{Department of Physics, University of Basel, 4056 Basel, Switzerland}

\author{W. D{\"u}r}
\affiliation{Institute for Theoretical Physics, University of Innsbruck, A-6020, Innsbruck, Austria}

\author{C. A. Muschik}
\affiliation{Institute for Theoretical Physics, University of Innsbruck, A-6020, Innsbruck, Austria}
\affiliation{Institute for Quantum Computing and Department of Physics and Astronomy,University of Waterloo, Waterloo, ON N2L 3G1, Canada}

\begin{abstract}
The vision to develop quantum networks entails multi-user applications, which require the generation of long-distance multi-party entangled states. The current rapid experimental progress in building prototype-networks calls for new design concepts to guide future developments. Here we describe an experimentally feasible scheme implementing a two-dimensional repeater network for robust distribution of three-party entangled states of GHZ type in the presence of excitation losses and detector dark counts --- the main sources of errors in real-world hardware. Our approach is based on atomic or solid state ensembles and employs built-in error filtering mechanisms peculiar to intrinsically two-dimensional networks. This allows us to overcome the performance limitation of conventional one-dimensional ensemble-based networks distributing multi-party entangled states and provides an efficient design for future experiments with a clear perspective in terms of scalability.

%Our approach is based on atomic or solid state ensembles -- scalable and experimentally available resources. Due to a built-in filtering mechanism, errors, arised in the elementary segments, pass thought the first quantum swapping operation with only  second order probability of their appearance, and further error propagation is suppressed. 
%Due to a built-in filtering mechanism, errors are passed thought the first quantum swapping operation with only second order probability of their appearance in the elementary segments, and further error propagation is suppressed. 
%We developed a novel diagrammatic technique for a numerical networks simulation, which shows that the proposed scheme is scalable and has much higher fidelity limit than a one-dimensional based network in presence of excitation losses -- the main source of errors for such a type of systems.

\end{abstract}
\maketitle

\bgroup\let\addcontentsline=\nocontentsline
\section*{Introduction}
The development of quantum networks holds the promise to realize quantum technologies such as  secure communication schemes~\cite{Bruss2000}, distributed quantum computing~\cite{Beals2013} and metrological applications~\cite{Komar14,Eldredge2016,Proctor2017,Ge2017}. This prospect 
%led to rapid experimental progress in this area and 
resulted in intense efforts to build prototype-networks with few nodes. The creation of long-range bipartite entangled states has been demonstrated~\cite{ExpDetTransfer,TracyReview,Chou2005,Yuan2008,Olmschenk2009,Lee2011,Krauter2013,Pfaff2014,Delteil2016,Hofmann2012,BellTest2} and multi-party entangled states shared between several remote parties can be realized in the near-future~\cite{SciNet}. 
The ability to distribute such multi-party entangled states over long-distances~\cite{Epping2016} will be an essential prerequisite for future quantum networks~\cite{KimbleReview,Ying2012,Pant2017,Siddhartha2018,Wehnereaam9288,Pirandola2016} consisting of large two-dimensional (2D) structures that allow for multi-user applications. This need and the current experimental possibilities raise the question which type of architecture can be used to distribute long-range multi-party entanglement in a practical and scalable fashion. We address this problem from an implementation-oriented point of view and propose a networking scheme that allows for a robust and efficient realization with current and near-future experimental means.
\begin{figure}
\includegraphics{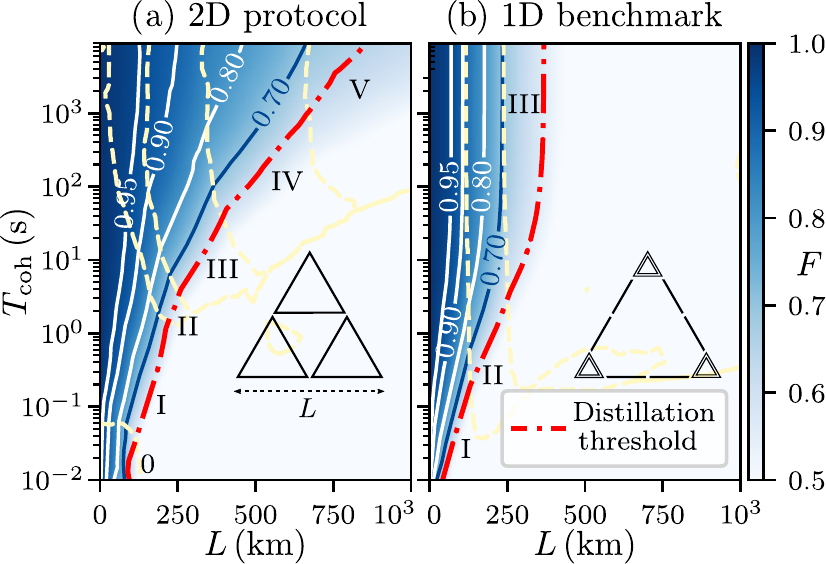}
\caption{Generation of long-distance GHZ states using (a) the 2D scheme shown in inset and Fig.~\ref{fig:2} and (b) the 1D benchmark (see inset and text). Both panels display the fidelity $F$ versus the memory excitation lifetime $T_{\text{coh}}$ and the covered distance $L$ for the parameters specified in Methods section. Red dash-dotted lines indicate the threshold beyond which entanglement distillation is impossible (see Supplemental Material, Sec.~D). Dashed lines separate network structures with a different optimal number of nesting levels (indicated with roman numbers).}
\label{fig:1}
\end{figure}

Future quantum networks will require different types of multi-party entangled states suitable for different tasks. We concentrate on distributing three-party GHZ states~\cite{GHZ1}, e.g. $\left(|001\rangle+|110\rangle\right)/\sqrt{2}$, between remote parties, that can be used for clock synchronization~\cite{Komar14}, quantum secret sharing~\cite{VotingAndSecretSharing,SecretSharing3GHZ}, quantum secret voting~\cite{Voting3GHZ} or for fundamental tests of nature~\cite{Mermin1990}.
A key challenge for distributing quantum states over long-distances is the fact that losses and decoherence scale exponentially in the distance~\cite{Takeoka2014,Pirandola2017}. To solve this problem, quantum repeaters~\cite{Briegel98,ReviewJiang,ReviewSangouard,Muralidharan2015} have been introduced, which however are inherently one-dimensional (1D) schemes aiming at generating a bipartite entangled state connecting two remote parties. Therefore, an intrinsically 2D network can be preferable for a multipartite entanglement distribution as it was shown~\cite{Wallnofer2016} to have a higher error threshold than a 1D counterpart under assumptions of a generic noise model and  full Bell state analysis (or a universal gate set) availability. 

In this work, we propose an implementation-oriented 2D repeater scheme that, in contrast to~\cite{Wallnofer2016}, (i) can be realized using a constrained set of quantum operations available in atomic or solid state ensembles and (ii) has a mechanism to mitigate an excitation loss error and detector dark counts --- major imperfections for this type of system.
%Here we propose and analyze a multi-user quantum repeater structure based on atomic or solid state ensembles. %Combining the restricted Bell state analysis and gates, experimentally available on the platforms, our scheme is able to %tolerate excitation loss errors which are relevant to the real-world hardware and conceptually different from the generic %noise model.
%
%Our scheme addresses not only a different task than the standard repeater scenario, but includes also an intrinsic error-filtering mechanism exploiting the multi-party entanglement of the generated states.
The resulting robust architecture is custom-tailored to implementations based on atomic or solid state ensembles, as these systems are promising for realizing long-lived quantum memories~\cite{deRiedmatten2015,QAP_review2010}. While atomic ensembles allow a limited set of operations for processing quantum information, record coherence lifetimes ($2T_\mathrm{coh}$)  up to six hours have been observed in rare-earth doped crystals~\cite{Zhong2015}. This makes them particularly promising for entanglement distribution over continental distances using repeater schemes which inevitably requires memory times corresponding to the duration needed for classical communication between remote nodes. 

Our scheme is inspired by the seminal DLCZ proposal for generating Bell pairs between two nodes~\cite{Duan2001}, and requires optical cavities with good cooperativities, linear optical elements, and photodetectors. GHZ states are first generated over moderate distances and then merged to form GHZ states connecting increasingly remote parties [see inset in Fig.~\ref{fig:1}(a) and Fig.~\ref{fig:2}(a)]. Similar to the DLCZ proposal, our protocol does not require a universal gate set or a full Bell-state analysis.
%In contrast to~\cite{Wallnofer2016}, our protocol does not require a universal gate set or a full Bell-state analysis. This is a property shared with the DLCZ proposal. 
%
However, unlike the DLCZ scheme, the new protocol suppresses the propagation of so-called vacuum and multi-excitation errors. These errors are resulted from excitation loss or detection of a dark count during the generation of the elementary state or merging operation and lead to the preparation of a state containing more/fewer excitations than expected. In the DLCZ scheme, the vacuum and multi-excitation errors freely propagate decreasing fidelity dramatically with the increase of nesting level and therefore making the scheme not scalable (\cite{ReviewSangouard} and Sec.~A1 of Supplemental Material). In comparison, our 2D scheme results in a truly scalable architecture. As shown in Fig.~\ref{fig:1} and explained below, this feature allows our repeater scheme to cover longer distances than networks based on combining conventional (one-dimensional) quantum links.
\begin{figure}
\includegraphics{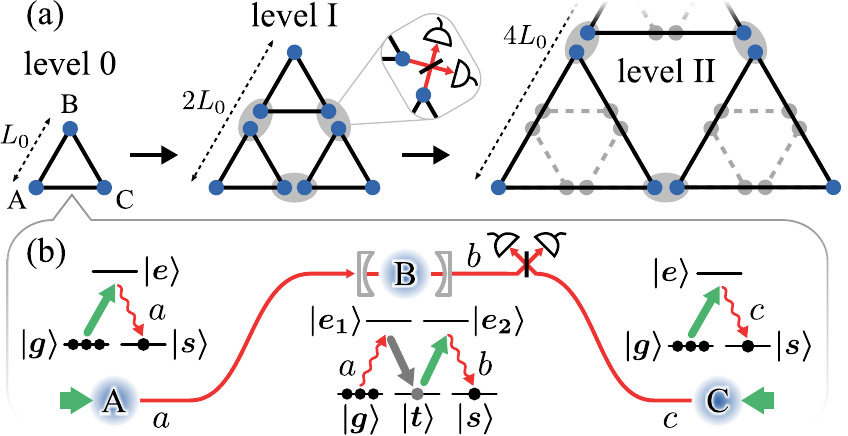}
\caption{2D repeater scheme. (a) Merging of GHZ states of three elementary segments spanning the length  $L_0$ into a GHZ state of a segment covering the length $2L_0$, which are further merged to a segment of the length $4L_0$. Blue circles represent ensembles, and triangles depict network segments.  The merging is realized by three swapping operations (see text) that are shown as gray ellipses. (b) Generation of the GHZ states at the elementary level. At nodes A and C, a \mbox{$\Lambda$-type} level structure is used allowing for standard DLCZ-type write-in and read-out operations~\cite{Duan2001}. Node B employs an ensemble with a \mbox{double-$\Lambda$} configuration placed in a cavity (see Supplemental Material, Sec.~G).}
\label{fig:2}
\end{figure}
%

%--------------------------------------------------------------------------------------------------------------------------------------------------------------------------------------------------------------------------
%--------------------------------------------------------------------------------------------------------------------------------------------------------------------------------------------------------------------------
%--------------------------------------------------------------------------------------------------------------------------------------------------------------------------------------------------------------------------
%--------------------------------------------------------------------------------------------------------------------------------------------------------------------------------------------------------------------------
%

\section*{Results}
\subsection*{Scheme}
%
% Scheme on a conceptual level
We propose a nested structure of quantum network~\cite{Wallnofer2016} consisting of segments in three-party entangled states of increasing size as shown in Fig.~\ref{fig:2}(a). The scheme requires the ability to generate GHZ states of elementary segments spanning a length $L_0$, and generalized entanglement swapping procedure to merge the GHZ states doubling the covered distance. We detail the generation and swapping operations separately in the next two paragraphs. 

% Generation of GHZ states on the lowest level
The generation of GHZ states at the elementary level is illustrated in Fig.~\ref{fig:2}(b). We consider two types of ensembles: (i) ensembles with a \mbox{$\Lambda$-level} scheme at nodes A and C, allowing for the efficient storage and read-out of photons~\cite{Duan2001,SimonJ2007,Gorshkov2007} and (ii) ensembles with a \mbox{double-$\Lambda$} configuration placed in a cavity at node B. Information is encoded in the absence or presence of a collective spin excitation in the ensembles, i.e. in the logical states $|0\rangle = |g_1g_2...g_N\rangle$ and $|1\rangle=\frac{1}{\sqrt{N}}\sum_{i=1}^{N} |g_1g_2...s_i...g_N\rangle$, where $N$ is the number of emitters per ensemble and $\ket{g}$, $\ket{s}$ denote ground state levels of the emitters [see Fig.~\ref{fig:2}(b)]. Ensembles A and C are driven by weak coherent laser pulses resulting in entangled states of the ensembles and the corresponding forward-scattered photons~\cite{Duan2001}, $\ket{0_{A(C)}0_{a(c)}}+\epsilon_{a(c)}\ket{1_{A(C)}1_{a(c)}}+\mathcal{O}(\epsilon_{a(c)}^2)$, $\epsilon_{a(c)}\ll1$ (the role of the parameter $\epsilon$ is explained in Sec.~B of Supplemental Material, our analysis includes also higher order terms). The capital and lowercase subscripts refer to the states of the ensemble and the corresponding emitted light field. Node B performs the gate operation 
$\ket{1_{a}0_B0_{b}}\rightarrow \ket{0_{a}1_B1_{b}}$,  $\ket{0_{a}0_B0_{b}}\rightarrow \ket{0_{a}0_B0_{b}}$.
%, where the subscript $a$~($b$) refers to the incoming~(outgoing) photonic mode. 
%This operation requires a mode mode matching condition~\cite{Vogell2017,Dilley2012,Mabuchi1997}).
The working mechanism and imperfections are detailed in Supplemental Material, Sec.~G. The light fields emitted from nodes B and C are synchronously directed to a swapping station equipped with a 50/50 beamsplitter and two single-photon detectors, as illustrated in Fig.~\ref{fig:2}(b). Conditioning on detecting a single photon allows for a probabilistic projection onto the state $\ket{\Psi_{\text{\tiny{GHZ}}}} = \left(|0_A0_B1_C\rangle+|1_A1_B0_C\rangle\right)/\sqrt{2}$. The path length fluctuation  in fibers AB and CB  can be solved by phase stabilization, similar to 1D repeater schemes~\cite{Sangouard2009}. For example, it can be achieved by exciting memories in A and C in a Sagnac interferometer configuration~\cite{Childress2005,Minar2008}, such that the excitation laser pulses for the two memories and the emitted photons travel the same path in a counter-propagating fashion.

% Merging of triangles
For merging three GHZ states, the states of ensembles at adjacent nodes are projected onto the one-excitation subspace. This operation is realized by reading out the atomic excitations~\cite{Duan2001,SimonJ2007,Gorshkov2007} and directing the emitted light fields to the swapping station described above [see also inset of Fig.~\ref{fig:2}(a)], where success is heralded by the detection of a single photon (see Sec.~A of Supplemental Material for details), otherwise the resulting state is discarded. 
%The error filtering mechanism associated with this merging procedure is specific for the considered is based on the observation that the most important experimental imperfections in the considered setting lead to errors loss of loss-type, as explained in the following.
%
%--------------------------------------------------------------------------------------------------------------------------------------------------------------------------------------------------------------------------
%--------------------------------------------------------------------------------------------------------------------------------------------------------------------------------------------------------------------------
%--------------------------------------------------------------------------------------------------------------------------------------------------------------------------------------------------------------------------
%--------------------------------------------------------------------------------------------------------------------------------------------------------------------------------------------------------------------------
%

\subsection*{Error filtering}
%
% Introduce the different types of errors
Apart from the major challenge to mitigate photon transmission losses, experimental limitations at the individual nodes have to be taken into account. The most important local error sources are the read-out inefficiency $v$ and imperfect photon detectors (with a detection inefficiency $f$ and a dark count probability $d$ during a photon pulse detection). 
Moreover, quantum states stored in the ensembles are degraded over time. Due to the encoding used here, the relevant physical mechanism --- dephasing of individual emitters --- leads to an effective decay of the stored collective excitations (Supplemental Material, Sec.~F). In the considered ensemble-based setting, the main imperfections are therefore errors of loss type and detector dark counts.

\begin{figure}
\includegraphics{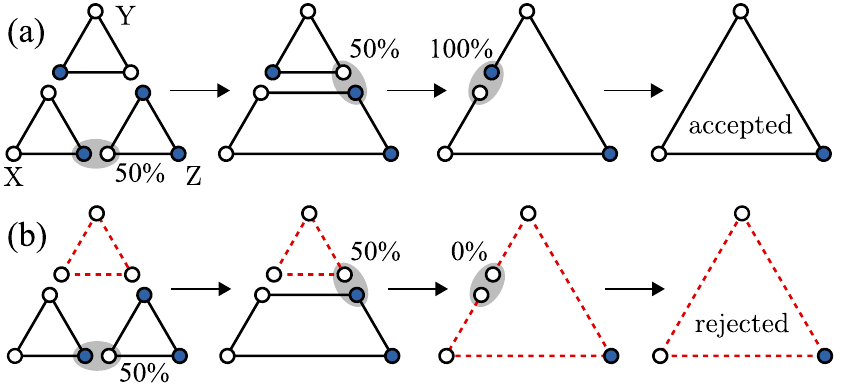}

\caption{Error filtering mechanism of the 2D repeater scheme. Solid (red dashed) lines represent quantum
(classical) correlations of the states of the nodes shown with circles. The states of the nodes of the same (different) colors are correlated (anti-correlated). Merging operations are shown with gray ellipses, the corresponding success probabilities (neglecting merging imperfections) are indicated nearby. (a) The no errors case: two successful probabilistic mergings of the three constituent GHZ states are followed by the third deterministic merging producing a perfect GHZ state $\left(\left\vert 0_{X}0_{Y}1_{Z}\right\rangle +\left\vert 1_{X}1_{Y}0_{Z}\right\rangle \right)\sqrt{2}$. (b) One loss type error or one multi-excitation error leading to preparation of a network segment in a wrong state (red triangle) can be deterministically filtered out.}

\label{Fig. 2} 
\end{figure}

The proposed 2D repeater protocol is designed to prevent the propagation of this kind of errors by employing an intrinsic redundancy of the native 2D network to filter out the errors in the merging process at each nesting level. The filtering mechanism, illustrated in Fig.~\ref{Fig. 2}, works as follows. At each nesting level one has to perform three merging operations to complete the entanglement generation between the three outermost nodes (X, Y, Z in Fig.~\ref{Fig. 2}). The third merging operation is redundant and it shows, if unsuccessful, that there is a lack or an excess of excitations in the generated state. This indicates an occurrence of either the loss type error or the multi-excitation error at previous stages of the protocol.
%Each probabilistic merging operation is thought to be successful upon detection of only one photon which is read out of two memories. In the case when no errors occur neither in the states nor in the merging operations, Fig.~\ref{Fig. 2}(a), probabilistic detection of a photon in each of two first  merging operations deterministically results to the last successful merging operation, after which a perfect constituent GHZ of the next nesting level is produced.  In contrast, for example, one loss error or one multi-excitation error in the memories participating in the merging operations, Fig.~\ref{Fig. 2}(b), deterministically leads to a failed merging, and thus the erroneous state is automatically rejected.
%One erroneous merging operation is filtered out in a similar way or leads to an erroneous state of a segment in the next nesting level where it could be again filtered out as shown in Fig.~\ref{Fig. 2}(b). 
In our protocol, the errors have, therefore, to conspire within \textit{one} nesting level to pass undetected. In contrast, in the 1D benchmark scheme the errors can conspire within \textit{all} levels of the bipartite state preparation to pass filtration in the last generation step. As a result, the 2D architecture exhibits a slower rate  of error probabilities growth with the increase of the network nesting level (see Supplemental Material, Sec.~A). %Thus, our protocol enables implementation of higher nesting level networks and, consequently, distribution of entangled states for longer distances. 

%In Supplemental Material, Sec.~A, 

%To analyze the suppression of the errors by the described filtering mechanism we consider a drop of fidelity, $F=\sqrt{\bra{\Psi_{\text{\tiny{GHZ}}}}\rho \ket{\Psi_{\text{\tiny{GHZ}}}}}$, of the states $\rho$ generated by the networks with respect to the target state $\ket{\Psi_{\text{\tiny{GHZ}}}}$ as a function of the networks nesting level $n$. 
%Since the error filtering mechanism by itself requires the redundant, third, imperfect swapping operation in each nesting level, here we focus only on the errors related to the imperfect swapping operations. Therefore, in this analysis, we considered the networks with ideal states of elementary segments (triangles and links for the 2D and 1D schemes, respectively) in the limit of perfect quantum memory. As a result, we obtained (see Supplemental Material, Sec.~A, ) that, despite the large amount of the imperfect swapping operations per level in the 2D network, in the limit of $d,~f,~v\ll1$, the infidelity scaling in the 2D repeater scheme is slower than in the 1D benchmark,

%\begin{align}
%1-F_{n}^{\text{1D}} & \sim\frac{15}{8}4^{n}\cdot(f+v)d,\label{eq:-44}\\
%1-F_{n}^{\text{2D}} & \sim\frac{3}{4}3^{n}\cdot(f+v)d.\label{eq:-45}
%\end{align}

However, the filtering is a passive operation, therefore, it is important to start the entanglement distribution with initial states of high fidelity. As detailed in Sec. A and B of Supplemental Material, the protocol for
generating entanglement at the elementary level and the relative orientation of GHZ states in the scheme is chosen such that the involved measurements allow for (i) identifying imperfect merging operations for ideal input states and (ii) filtering out imperfect states for ideal merging operations. The initial states of high fidelity at the elementary level and the subsequent error filtering at each nesting level result in the scheme robust against the vacuum and multi-excitation errors.

%An important feature of the scheme is its robustness to so-called multi-excitation errors that arise in DLCZ-type schemes if coherent input fields are used and that are typically a major problem for 1D schemes.
%
%--------------------------------------------------------------------------------------------------------------------------------------------------------------------------------------------------------------------------
%--------------------------------------------------------------------------------------------------------------------------------------------------------------------------------------------------------------------------
%--------------------------------------------------------------------------------------------------------------------------------------------------------------------------------------------------------------------------

\subsection*{Key features}
In the following, we highlight some of the main properties of our scheme and summarize the effect of limited memory coherence times in the presence of experimental imperfections. Figure~\ref{fig:1}(a) shows the fidelity of the generated state $\rho$ with respect to the target state $\ket{\Psi_{\text{\tiny{GHZ}}}}$, $F=\sqrt{\bra{\Psi_{\text{\tiny{GHZ}}}}\rho \ket{\Psi_{\text{\tiny{GHZ}}}}}$, for the proposed architecture. 
In Fig.~\ref{fig:1}(b), for comparison, we introduce a fidelity benchmark representing the performance of a 1D scheme based on long-distance bipartite entanglement distribution. The 1D benchmark uses conventional ensemble-based repeater scheme which creates three long-distance Bell pairs to be subsequently merged to generate the desired target state. The benchmark estimates an upper bound of the fidelity as ideal local GHZ states [shown with double lined triangles in the inset in Fig.~\ref{fig:1}(b)] are used for the final merging. In contrast to the 2D approach, which relies on multi-party entanglement at all network levels, the 1D scheme exploits the distribution of long-distance Bell pairs and involves the error filtering mechanism only in the end~(Methods section).

%To compare our approach to schemes based on long-distance bipartite entanglement distribution, Fig.~\ref{fig:1}(b) shows the fidelity achievable with a 1D benchmark -- conventional ensemble-based repeater scheme which creates three long-distance Bell pairs to be subsequently merged to generate the desired target state using ideal local GHZ states [see inset in Fig.~\ref{fig:1}(b)]. In contrast to the 2D approach, which relies on multi-party entanglement at all network levels, the 1D scheme relies on the distribution of long-distance Bell pairs and involves the error filtering mechanism only in the end~( Supplemental Material, Sec.~A2).
%
%Both plots in Fig.~\ref{fig:1} display the fidelity $F(T_{\text{coh}},L)$ as a function of the memory coherence time $T_{\text{coh}}$ and the covered distance $L$.
%
The most striking feature of Fig.~\ref{fig:1} is the fact that the 2D protocol can distribute GHZ states over increasingly large distances by increasing the memory coherence time (or alternatively, by using so-called multiplexing approaches~\cite{Multiplexing,Simon2007,Bonarota2011,Abruzzo2014,vanDam2017} in which several quantum memories or memory modes are used in parallel to compensate for limited coherence times). 
An increase of $T_{\text{coh}}$ significantly extends the distance that can be covered, since our protocol profits from using higher nesting levels for larger distances. This property makes the scheme practical and scalable. This is not the case for the 1D benchmark based on a regular DLCZ scheme, which is hampered by fundamental difficulties associated with the creation of long-range bipartite links (see Supplemental Material, Sec.~A3 and Sec.~C for details). 
%Therefore, the 1D strategy reaches the ``performance wall'', a distance beyond which investments in resources or an increase of $T_{\text{coh}}$ do not lead to an increase in fidelity [vertical fidelity-lines in Fig.~\ref{fig:1}(b)], much sooner than the 2D scheme.
%Therefore, the 1D strategy reaches the infinite-memory limit (at which the fidelity-lines in Fig.~\ref{fig:1}(b) become vertical and an increase in $T_{\text{coh}}$ does not lead to an increase in $L$) at much shorter distances than the 2D scheme.
Therefore, the 1D strategy reaches its performance limit %(at which the fidelity-lines in Fig.~\ref{fig:1}(b) become vertical and an increase in $T_{\text{coh}}$ does not lead to an increase in $L$) 
at much shorter distances than the 2D scheme as can be seen in Fig.~\ref{fig:1}(b) in comparison with Fig.~\ref{fig:1}(a).

To further characterize the potential of our approach, we analyze the case of  infinite memory coherence time. Figure~\ref{fig:3} shows the fidelity  achievable by a network for which the number of nesting levels is optimal for a given distance in the presence of typical experimental imperfections for $T_{\text{coh}}\rightarrow \infty$. 
A network with more nesting levels divides the target distance into shorter segments what reduces the probability to lose photons during transmission at the cost of additional errors generated by the repeater scheme itself. %These errors are resulted from additional imperfect merging operations, but also from  stacking of errors presented in the imperfect segments whose number grows with the nesting level increase.  
For each target distance, there is an ideal number of nesting levels beyond which the addition of a repeater station becomes undesirable. 
The comparison of the 2D approach (black line) with the 1D benchmark (green line) for generating three-party GHZ states shows that the latter suffers from severe limitations in the achievable fidelities for large distances even for perfect quantum memories. This is the result of the faster errors accumulation with the growing network size in 1D scheme which lacks the filtering mechanism of the native 2D protocol, as discussed in the previous section.
\begin{figure}
\includegraphics{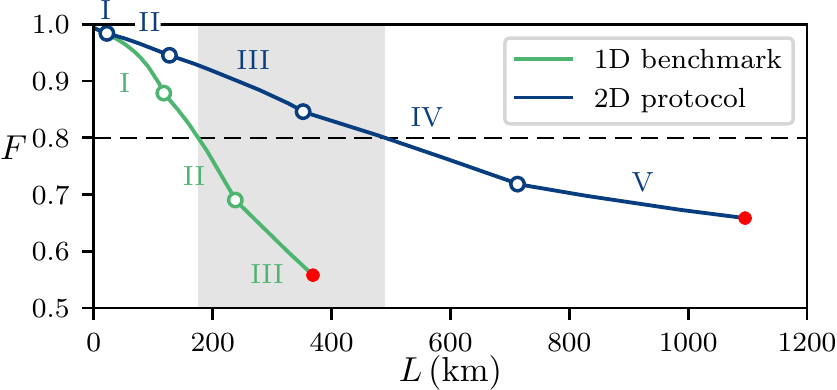}
\caption{Fidelity $F$ for the proposed 2D scheme and the 1D benchmark (see text) in the limit of infinite memory coherence time, as a function of the covered distance $L$. This plot corresponds to a cut of Fig.~\ref{fig:1} at $T_{\text{coh}}\rightarrow\infty$. The gray area illustrates that for large distances, certain target fidelities can only be reached by the 2D  scheme (the desired target fidelity depends on the application, in this example $F_{\text{target}}=0.8$). Roman numbers indicate the optimal number of nesting levels for the network. Transitions between the optimal  numbers of levels are shown by circles. Red dots indicate the distance  beyond which entanglement distillation is impossible.}
\label{fig:3}
\end{figure}
%
%The comparison of the 2D approach (black line) with the 1D benchmark (green line) for generating three-party GHZ states shows that the latter suffers from severe limitations in the achievable fidelities for large distances even for perfect quantum memories. %Our approach mitigates this problem and allows for high target fidelities of long-distance GHZ state distribution. 

%~(\ref{eq:-44})

Here we emphasize that although the 1D benchmark uses fewer resources (memory cells) than our 2D scheme (see Supplemental Material Sec.~A2) its performance is saturated at much shorter distances and can not be improved by a simple increase of resources invested into multiplexing. The presented 2D architecture allows us to convert the extra resources into longer entanglement distribution distances with higher target fidelities. Moreover, the additional memory cells improve coverage of intrinsically 2D networks by allowing the multi-party entanglement generation between arbitrary nodes of the network (see Supplemental Material Sec.~A1), while 1D based schemes entangle only the outermost ensembles.

A detailed analysis of the 2D protocol performance, including an assessment of the achievable rates, is provided in Supplemental Material (Sec.~A3 and Sec.~C). In the following, we discuss means to further improve the performance by temporal filtering. %and multiplexing~\cite{Multiplexing} 
% in detail.%and discuss the experimental realization in a realistic setting.
%
%--------------------------------------------------------------------------------------------------------------------------------------------------------------------------------------------------------------------------
%--------------------------------------------------------------------------------------------------------------------------------------------------------------------------------------------------------------------------
%--------------------------------------------------------------------------------------------------------------------------------------------------------------------------------------------------------------------------

\subsection*{Temporal filtering protocol}
Limited coherence times of quantum memories are generally an important restricting factor for quantum repeaters. 
%For example, for the specific set of parameters used in Fig.~\ref{fig:1}, the proposed repeater scheme starts to outperform direct transmission (corresponding to the elementary level) roughly at $t_{\text{mem}} = 300$ms. 
To mitigate their effect, we introduce a temporal filtering mechanism by defining a time window $\tau$, after which quantum states are discarded. During the probabilistic generation and merging procedure shown in Fig.~\ref{fig:2}, ensembles storing excitations longer than $\tau$ are reset to their ground state, such that the entanglement generation process involving these nodes starts anew. As a result, the influence of decoherence is decreased at the expense of a reduced rate. The time $\tau$ can be changed dynamically by the control software without changing the hardware in a quantum network. Due to this added flexibility, different types of applications --- requiring either high fidelities or high rates --- can be accommodated. The corresponding tradeoff has been calculated semi-analytically (as described below) and is shown in Fig.~\ref{fig:4}.
%We note that in the limit $tau \ll 1$, the temporal filtering corresponds to waiting for the simultaneous generation of all network pieces. This is the limit of so called quantum relay.  For short memories the filtering is a way to enhance a quantum relay protocol
%
%
\begin{figure}
\includegraphics{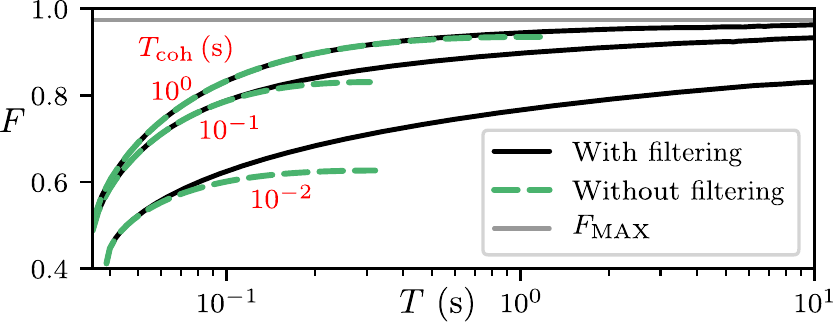}
\caption{Temporal filtering for a 2D network with nesting level I, i.e. 9 ensembles [see Fig.~\ref{fig:2}(b)] for the parameters specified in Methods section.
The fidelity $F$ is shown versus the entanglement generation time $T$ for a target distance $L=50$ km and different memory times $T_{\text{coh}}=\{10^{-2}, 10^{-1},1\}$s. Black solid (dashed green) lines represent results with (without) temporal filtering. The gray solid line indicates the fidelity in the limit of infinite quantum memory times.}
\label{fig:4}
\end{figure}
%
%--------------------------------------------------------------------------------------------------------------------------------------------------------------------------------------------------------------------------
%--------------------------------------------------------------------------------------------------------------------------------------------------------------------------------------------------------------------------
%--------------------------------------------------------------------------------------------------------------------------------------------------------------------------------------------------------------------------
%

\subsection*{Diagrammatic technique} We used a numerical Monte Carlo algorithm and developed a new semi-analytical technique to analyze the performance of our scheme in the presence of realistic imperfections. The Monte Carlo method is very flexible and simulates the full repeater protocol step by step \cite{Cody2016, VanMeter2017}. The strength of this exact simulation is also its weakness --- the runtime is proportional to the entanglement generation time in the quantum network, growing quickly with the network's scale.

The analytical method overcomes this difficulty while still incorporating all relevant error sources, including finite memory coherence times. It takes time delays due to classical communication into account and can be used to analyze a large class of repeater schemes, including 1D protocols. The main idea is to determine the density matrix distribution $\varrho(t)$ for the ensemble of states generated by the network at time $t$. More precisely, we obtain the Laplace image of this distribution, $\tilde\varrho(s)$, which fully describes the statistics of the network and allows one to infer the average generated state $\rho\equiv\int_0^{\infty}\varrho(t)dt=\tilde\varrho(s)\big|_{s=0}$, corresponding generation time $T\equiv\int_0^{\infty}t\,{\rm Tr}\varrho(t)dt=-\frac{d}{ds}{\rm Tr}\tilde\varrho(s)\big|_{s=0}$, and other relevant statistical quantities.

We assume that the probability $p_1$ to generate an entangled state of an elementary segment is small for each time step $\Delta t$, such that the continuous probability density $p(t)\equiv\nu e^{-\nu t}$ with the rate $\nu\equiv p_1/\Delta t$ can be introduced. As an example, we consider a linear network consisting of two links, where the density matrix distribution for generating the $\alpha^{\text{th}}$ link in a state $\rho_\alpha$ before the $\beta^{\text{th}}$ link has been generated at time $t$ in a state $\rho_\beta$ is given by $\varrho_{\alpha\beta}(t)=p_\beta(t)\int_0^{t} dt'\,p_\alpha(t')e^{(t-t')\mathcal{L}_\alpha}\rho_1\otimes\rho_2$ with ${\alpha\beta}\in\{{12,21}\}$. Here we sum over all intermediate times $t'$ of the $\alpha^{\text{th}}$ link creation. The degradation of the link during the waiting time $t-t'$ due to finite memory life times is taken into account using the superoperator $\mathcal{L}_\alpha$. The corresponding Laplace image of the distribution is given by $\tilde\varrho_{\alpha\beta}(s) = \nu_\alpha\nu_\beta/[(s' + \nu_\alpha)(s'-\mathcal{L}_\alpha)]\cdot\rho_1\otimes\rho_2 \big|_{s'=s+\nu_\beta}$.

The probability to generate and successfully merge states of two segments during the time window $[t,t+dt)$ is given by ${\rm Tr}\,\mathcal{M}\varrho_{\alpha\beta}(t)dt$, with $\mathcal{M}$ the merging superoperator. The summation over all possible combinations of unsuccessful mergings leading to the generation of an entangled state results in a sum of multiple convolutions in the time domain. In the Laplace domain, the sum converges to
\begin{equation}
\tilde\varrho\left(s\right)=\frac{\sum_{\alpha\beta}\mathcal{M}\tilde\varrho_{\alpha\beta}(s)} {1-\sum_{\alpha\beta}{\rm Tr} \left(\mathbb{\mathcal{I}}-\mathcal{M}\right)\tilde\varrho_{\alpha\beta}(s) }\ ,\label{eq:rho_s}
\end{equation}
with $\sum_{\alpha\beta}$ the sum over all possible orders in which links are generated and $\mathcal{I}$ the unit superoperator. The Laplace image~\eqref{eq:rho_s} is used to find the average density matrix $\rho'$ and generation time $T'$ for the given nesting level of the network, as described above. To address the next network level, we apply the approximation that the segments are generated time-independently in the state $\rho'$ with the rate $\nu'=1/T'$. In a recursive procedure, the state $\rho$ and generation time $T$ of an arbitrary network level can be found. As  detailed in Sec.~E of Supplemental Material, we introduce diagrams to conveniently deal with probabilistic processes in networks of arbitrary complexity. 
%
%--------------------------------------------------------------------------------------------------------------------------------------------------------------------------------------------------------------------------
%--------------------------------------------------------------------------------------------------------------------------------------------------------------------------------------------------------------------------
%--------------------------------------------------------------------------------------------------------------------------------------------------------------------------------------------------------------------------

\section*{Discussion}
We proposed a scalable 2D architecture for generating long-distance multi-party entanglement and provided a full performance analysis with emphasis on the effect of finite memory lifetimes. Covering increasingly longer distances as shown in Fig.~\ref{fig:1} can be achieved either using memories with long coherence times or using multiplexing approaches~\cite{Multiplexing,Simon2007,Bonarota2011,Abruzzo2014,vanDam2017}, in which several memory-modes are used in parallel. The presented scheme provides a flexible structure for creating GHZ states between arbitrary nodes of the 2D network, not only between the outermost ensembles. The logical topology of the network [shown in Fig.~\ref{fig:2}(a)] is thereby not identical to the topology of the required fiber links, which can easily be adjusted to accommodate urban constraints.
We are here primarily interested in metropolitan distances and applications requiring only moderate bit-rates. Important examples include secret voting and the protection of classified information that requires several parties for decryption~\cite{VotingAndSecretSharing,SecretSharing3GHZ}. 

The presented scheme can be modified to work without cavities at the expense of using a larger number of ensembles (see Supplemental Material, Sec. H). 
This modification uses  polarization-type qubits~\cite{Duan2001,ReviewSangouard} and two-click conditioning~\cite{Sangouard2008}, enhancing its resistance with respect to fiber-length fluctuations.
It will be interesting to develop similarly robust schemes for other platforms such as trapped ions~\cite{TracyReview,Duan2010}, in which a universal gate set can be implemented. In this case, more complicated 2D repeater protocols involving quantum error correction~\cite{ReviewJiang} instead of error filtering could be envisaged.
%and encoded 2D architectures inspired by so-called repeaters of second and third generation could be studied.
%

\section*{Methods}
\subsection*{Parameters of the simulated network}
The results of simulations presented in Figs.~\ref{fig:1},~\ref{fig:3} and \ref{fig:4} are obtained with the following parameters of the network: detection inefficiency $f=0.05$, read-out inefficiencies $v=0.05$, fiber attenuation length $L_\mathrm{att}=22$km, gate efficiency (see text) $\eta=0.6$, signal length $l= 10^{-4}$s, and dark count probability in the signal measurement operation $d=10\mathrm{Hz}\cdot10^{-4}\mathrm{s}=0.001$.

\subsection*{One-dimensional benchmark scheme}\label{Sec.1Dscheme}

\begin{figure}

\includegraphics{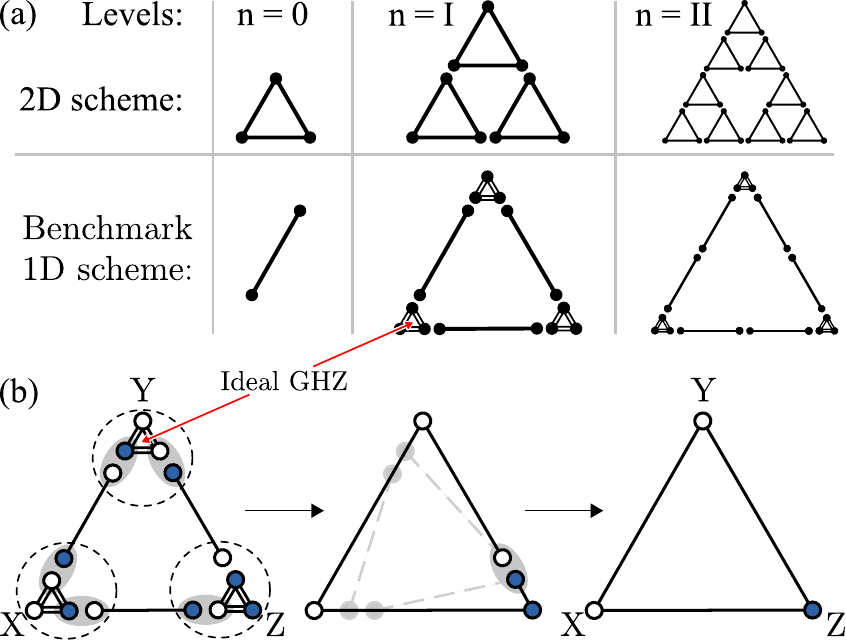}

\caption{One-dimensional benchmark scheme. (a) Comparison of the 2D scheme and the 1D benchmark scheme at different nesting levels. Dots represent quantum memories. Dots connected by lines represent elementary segments of the networks. Triangles with double lines indicate ideal GHZ states. (b) The last step of generating $\left\vert \Psi_{\rm GHZ}\right\rangle $ in the 1D benchmark scheme. The pictorial representation of states is explained in Fig.~\ref{Fig. 2}. The three distant parties X, Y, and Z hold perfect tripartite GHZ states, represented by triangles. The final filtering step described in the text is depicted in two stages, shown by arrows. Gray dashed lines and circles indicate the actual successive merging pattern, during which the long-range quantum links are first merged with the local GHZ states and then with each other. }
\label{Fig. 1} 
\end{figure}

Here we present an alternative, one-dimensional (1D), repeater approach for distribution of multi-partite GHZ state, which provides us with the fidelity benchmark used in the present work. This 1D benchmark scheme is based on the generation of long-range bipartite quantum states that are created using regular DLCZ quantum repeaters. A comparison of various nesting levels structures for the 1D benchmark scheme and the propose 2D scheme is presented in Fig.~\ref{Fig. 1}(a). 

While two links are enough to distribute entanglement between three parties, we equip the 1D scheme with the third link, which is used to implement error filtering procedure in the final step, similar to the original DLCZ protocol, as explained below. To ensure the robustness of the benchmark we assume that the distant parties X, Y, and Z of the 1D scheme have access to ideal GHZ states that can be generated on demand. This provides an upper bound on the performance of 1D schemes of DLCZ type for distribution of tripartite GHZ states. Therefore, the corresponding fidelity of the 1D scheme serves as a benchmark for our 2D scheme.

Figure~\ref{Fig. 1}(b) shows the final merging step of the 1D scheme acting as an error filtering operation. In this step, five swapping operations are followed by the last one, which is deterministic in the ideal case. A failure of the last swapping operation indicates an erroneous state. More details about the 1D benchmark scheme are presented in  Sec.~A of Supplemental Material.

\section*{DATA AVAILABILITY}
The data that support the findings of this study are available from the corresponding author upon reasonable request.

\begin{acknowledgements}
We thank M.  Afzelius and P. Jobez for fruitful discussions on solid-state ensemble-based implementations and J. Walln{\"o}fer for his input on 2D architectures.
Research was sponsored by the Swiss National Foundation (SNSF) through grant number PP00P2-179109, by the Austrian Science Fund (FWF): P28000-N27, P30937-N27, and by the Army Research Laboratory and was accomplished under Cooperative Agreement Number W911NF-15-2-0060. The views and conclusions contained in this document are those of the authors and should not be interpreted as representing the official policies, either expressed or implied, of the Army Research Laboratory or the U.S. Government. The U.S. Government is authorized to reproduce and distribute reprints for Government purposes notwithstanding any copyright notation herein. This research was undertaken thanks in part to funding from CFREF.

\end{acknowledgements}

\textit{Note added.} Recently we have become aware of~\cite{Santra2019} discussing a similar temporal filtering technique.

\section*{AUTHOR CONTRIBUTIONS}
All authors contributed equally to the current paper.

\section*{ADDITIONAL INFORMATION}

\paragraph*{Competing interests:}
The authors declare no competing interests.

\egroup

\newpage

\appendix
\part*{Supplementary material}

The supplementary material contains eight sections. In Sec.~\ref{sec:Error-filtering-mechanism}, we explain the built-in error filtering mechanism of the two-dimensional (2D)
quantum repeater scheme put forward in the main text. We introduce the 1D repeater scheme used in the main text as a benchmark, and analyze performance of the 2D scheme. Sec.~\ref{sec:Elementary-triangle-generation} discusses the generation of the elementary GHZ states in detail. In Sec.~\ref{sec:Distribution-Time}, we provide results for the average generation time of long-range GHZ states using the proposed repeater architecture, and in Sec.~\ref{sec:Distillability-of-a}, we briefly review the distillability criterion used in the main text~\cite{Dur1999}. In Sec.~\ref{sec:Methods} we explain the numerical methods that have been used to study the performance of our scheme. In Sec.~\ref{sec:Memory-decoherence}, the model used to describe decoherence in quantum memories is set forth. In Sec.~\ref{sec:Double--scheme-ensemble}, we provide details on the nodes involving a double-$\Lambda$ level scheme. In Sec.~\ref{sec:Cavityless-scheme-of}, we describe an alternative version of our quantum repeater scheme that does not require cavities, but instead relies on larger number of resources and longer memory coherence time.

\tableofcontents

\section{Error filtering mechanism\label{sec:Error-filtering-mechanism}}

In this section we describe the basic working principles of the two-dimensional~(2D) quantum repeater scheme put forward in the paper and explain its built-in error filtering
mechanism. To demonstrate the effectiveness of this architecture, we compare our new 2D repeater scheme with a benchmark based on long-range quantum links that are generated using one-dimensional~(1D) repeaters. In particular, we analyze the scaling of the respective achievable fidelities with increasing repeater nesting levels.

\begin{figure}
\includegraphics{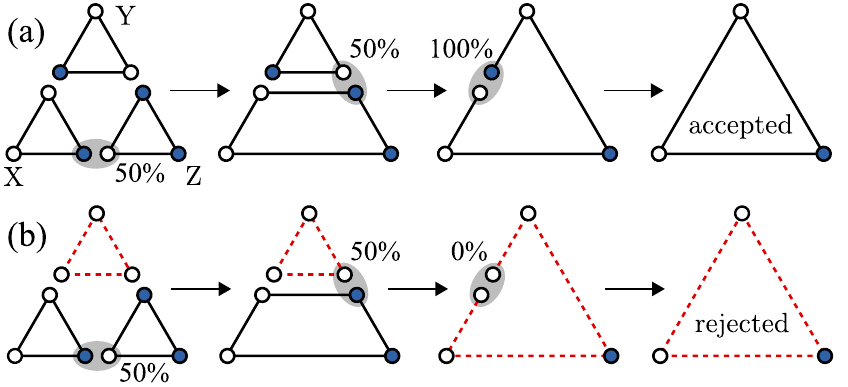}

\caption{Error filtering mechanism of the 2D repeater scheme. Solid (red dashed) lines represent quantum
(classical) correlations of the states of the nodes shown with circles. The states of the nodes of the same (different) colors are correlated (anti-correlated). Merging operations are shown with gray ellipses, the corresponding success probabilities (neglecting merging imperfections) are indicated nearby. (a) The no errors case: two successful probabilistic mergings of the three constituent GHZ states are followed by the third deterministic merging producing a perfect GHZ state $\left(\left\vert 0_{X}0_{Y}1_{Z}\right\rangle +\left\vert 1_{X}1_{Y}0_{Z}\right\rangle \right)\sqrt{2}$. (b) One loss type error or one multi-excitation error leading to preparation of a network segment in a wrong state (red triangle) can be deterministically filtered out.}

\label{Fig. 2_SM} 
\end{figure}

\subsection{Two-dimensional quantum repeater scheme}\label{Sec.2Dscheme}

In the main text, we propose a 2D repeater architecture based on~\cite{Wallnofer2016} and consider 
its application for the distribution of long-range tripartite GHZ states
\begin{eqnarray}
\label{Eq.GHZ} 
\left\vert \Psi_{\rm GHZ}\right\rangle =\frac{1}{\sqrt{2}}\left(\left\vert 0_{X}0_{Y}1_{Z}\right\rangle +\left\vert 1_{X}1_{Y}0_{Z}\right\rangle \right),
\end{eqnarray}
shared between three parties X, Y, and Z. Our 2D scheme is inspired by the original (one-dimensional) DLCZ proposal~\cite{Duan2001} and utilizes similar ingredients: linear optical elements, photodetectors and atomic or solid state ensembles. As an additional element we introduce an optical cavity with a good cooperativity, which is used for the generation of GHZ states at the elementary level of the repeater protocol (see Sec.~\ref{sec:Elementary-triangle-generation}). We note that the cavity can be dispensed with at the expense of using a large number of ensembles, as explained in Sec.~\ref{sec:Cavityless-scheme-of}.

\begin{figure}
\includegraphics{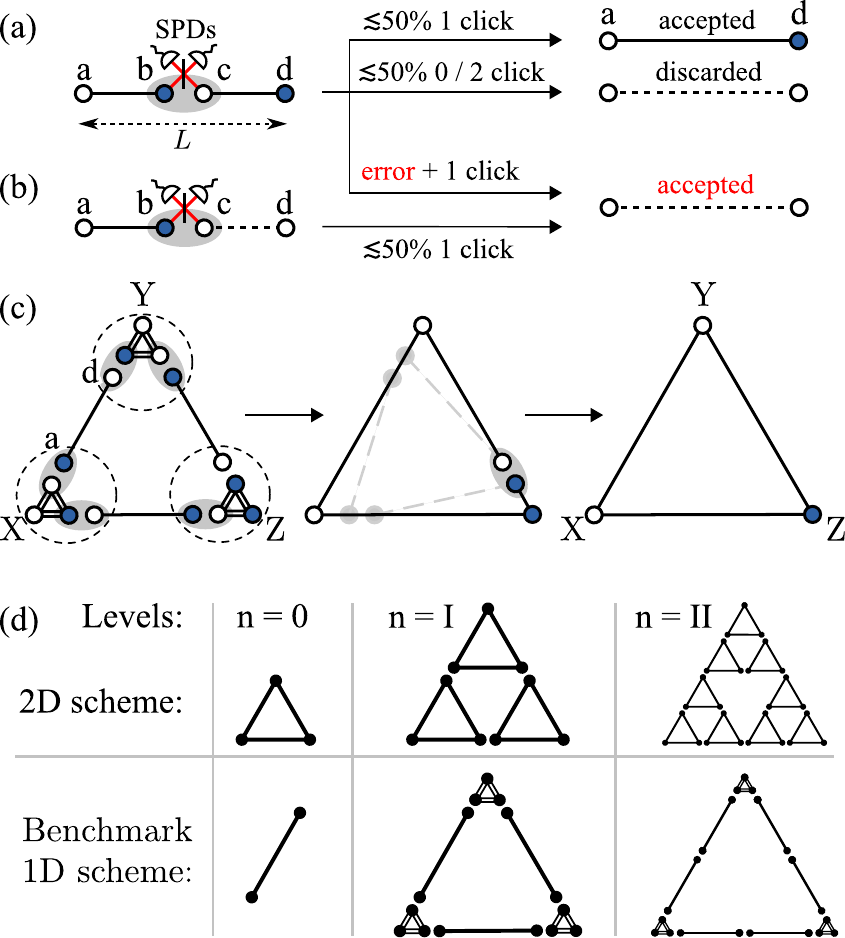}

\caption{One-dimensional benchmark scheme. (a) Bell state distribution via entanglement swapping. The pictorial representation of states is explained in Fig.~\ref{Fig. 2_SM}, red lines represent light modes, and  half circles denote single-photon detectors (SPDs). The measurement outcomes, their probabilities and the resulting states are indicated.
(b) Propagation of an error during entanglement swapping. (c) The last step of generating $\left\vert \Psi_{\rm GHZ}\right\rangle $ in the 1D benchmark scheme. The three distant parties X, Y, and Z hold perfect tripartite GHZ states, represented by triangles with double lines. The final filtering step described in the text is depicted in two stages, shown by arrows. Gray dashed lines and circles indicate the actual successive merging pattern, during which the long-range quantum links are first merged with the local GHZ states and then with each other. (d) Comparison of the 2D scheme and the 1D benchmark scheme at different nesting levels. Dots represent quantum memories. Dots connected by lines represent elementary segments of the networks.}
\label{Fig. 1_SM} 
\end{figure}

As the original DLCZ protocol, our 2D repeater scheme relies on probabilistic merging operations (explained below). Due to the probabilistic character of the entanglement creation and merging procedures, it is advantageous to employ a nested structure as shown in Fig.~2a in the main text. Figure~\ref{Fig. 2_SM}(a) shows a single nesting level of our 2D scheme, in which three network segments in tripartite GHZ states are merged by three entanglement swapping operations, thereby extending the distance over which entanglement is shared. The merging operation for bipartite states is illustrated in Fig.~\ref{Fig. 1_SM}(a): the states of two quantum memories are mapped to light fields which subsequently interfere at a balanced beamsplitter. Each of the two output ports of the beamsplitter is equipped with a single-photon detector (SPD), and the detection of a single photon in the measured light fields projects the joint system onto the desired entangled state.

Imperfections in realistic setups, such as fiber losses, memory read-out losses, memory decay, detector inefficiencies and dark counts --- which we consider in our analysis --- lead to errors
in the states distributed by quantum repeaters. Figure~\ref{Fig. 1_SM}(b) illustrates how loss errors or multi-excitation errors (see Sec.~\ref{subsec:Analysis}) can freely propagate through the merging of 1D repeater links by passing unnoticed through the entanglement swapping operations. The original DLCZ repeater protocol entails therefore an error filtering procedure as final step of its implementation~\cite{Duan2001}. However, this filtering step at the end of the protocol is ineffective against errors accumulated during the distribution of the quantum states~\cite{ReviewSangouard}. Therefore, our 2D repeater scheme uses a filtering mechanism at each nesting level of its scheme as illustrated in Fig.~\ref{Fig. 2_SM}(b). For each nesting level, two entanglement swapping operations probabilistically merge two tripartite GHZ states into a five-partite GHZ state. In the ideal case, the third entanglement swapping operation deterministically prepares the desired state
as the measurement of two anti-correlated qubits always results in a single detector click. An unsuccessful third merging operation during which no or more than one detector click is obtained indicates that there has been an error and leads to a rejection of the produced state. 

The filtering mechanism of the 2D scheme adds an extra swapping operation per each nesting level leading to a faster growth of the total number of swapping operations compared to a 1D benchmark scheme presented in the next section. Due to inevitable imperfections of the swapping process the probability of generating errors grows together with the number of swapping operations. Nevertheless, in Sec.~\ref{subsec:Performance}, we will show that the fidelity drops slower with the increase of nesting level $n$ for the 2D scheme than for the 1D benchmark. This means that the filtering mechanism eliminates more errors than it adds.

\begin{figure}
\includegraphics{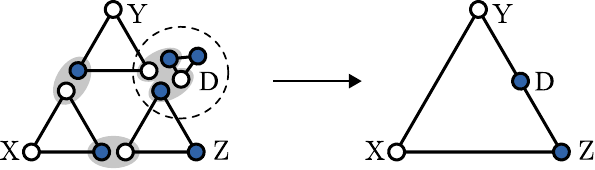}

\caption{Entanglement distribution between intermediate
nodes of 2D network: a local tripartite GHZ state is merged with two network
segments using two standard entanglement swapping operations.}
\label{Fig. 1_SM-1} 
\end{figure}

Apart from error mitigation, the  intermediate nodes of the 2D architecture serves the important purpose of allowing for the generation of GHZ states between arbitrary communication partners in a 2D network as shown in Fig.~\ref{Fig. 1_SM-1}.

\subsection{One-dimensional benchmark scheme}\label{Sec.1Dscheme_SM}

Here we present an alternative repeater approach for distribution of multi-partite GHZ state, shown in Fig.~\ref{Fig. 1_SM}(c), which provides us with a fidelity benchmark used in the main text. This one-dimensional (1D) benchmark scheme is based on the generation of long-range bipartite quantum states that are created using regular DLCZ quantum repeaters. While two links are enough to distribute entanglement between three parties, we equip the 1D scheme with the third link, which is used to implement error filtering procedure in the final step, similar to the original DLCZ protocol, as explained below. To provide a clear comparison and to reduce the discussion to the errors that occur within the different types of repeaters, we consider the 1D scheme assuming that the distant parties X, Y, and Z have access to ideal GHZ states that can be generated on demand. This provides an upper bound on the performance of 1D schemes of DLCZ type for distribution of tripartite GHZ states. Therefore, the corresponding fidelity of the 1D scheme serves as a benchmark for our 2D scheme.

Figure~\ref{Fig. 1_SM}(c) shows the final merging step of 1D scheme acting as an error filtering operation. In this step  five swapping operations are followed by the last one, which is deterministic in the ideal case. A failure of the last swapping operation indicates an erroneous state. Due the probabilistic nature of entanglement swapping operations, it is statistically faster to perform this final merging step  in two successive stages. In the first stage each link is merged with a corresponding ideal GHZ state, and in the second stage the resulted tripartite states are merged together as indicated in Fig.~\ref{Fig. 1_SM}(c) by gray dashed lines. A comparison of various nesting levels structures for the 2D scheme and the 1D benchmark scheme is presented in Fig.~\ref{Fig. 1_SM}(d).

Prior to the final filtering step of the 1D benchmark approach, errors propagate through the whole process during which the long-range bipartite entangled quantum links are established, as noted in the previous section [see Fig.~\ref{Fig. 1_SM}(b)]. Combinations of different types of errors often render them undetectable for the final filtering step. By comparison, the filtering at each nesting level used in the 2D approach reduces probability to obtain an undetectable combination of errors at the expense of a more rapid growth of resources. 

The required resources as a function of the network linear size $L_n=2^n$ with $n$ the network nesting level can be estimated as follows. The 2D strategy requires $3^{n+1}=3L_n^{\text{Log}_23}\approx3L_n^{1.6}$ memory cells as opposed to the 1D benchmark scheme using $3\cdot2^{n}+9=3L_n+9$, including the ideal local GHZ states at the outermost nodes~\footnote{
At the elementary level $n=0$ of the two repeater schemes, the basic
segments consist of tripartite GHZ states for the 2D strategy and bipartite
links (Bell states) for the 1D protocol. If the latter consists of $n$ levels in total, three bipartite long-range links generated by DLCZ repeaters involving $n-1$ levels are merged into a  
tripartite GHZ state in the last step, as shown in Fig.~\ref{Fig. 1_SM}(c). This final merging
step represents the $n^{\text{th}}$ nesting level in the 1D case.}. 
%Figure~1 in the main text shows that such an error filtration strategy leads to a significantly improved fidelity scaling.

\subsection{Performance of the two-dimensional scheme\label{subsec:Performance}}

In this subsection we numerically and analytically analyze the fidelity
scaling of the 2D repeater and the 1D benchmark schemes discussed above. More specifically, we calculate
the drop of the fidelity $1-F_{n}^{\text{1D(2D)}}$ as a function of the
total number of imperfect swapping operations required for the $n^{\text{th}}$
nesting level,
\begin{gather}
\begin{split}
N_{n}^{\text{1D}} & =3\left(2^{n-1}-1\right)+6,\\
%N_{n}^{\text{1D}} & =\frac{3}{2}\left(2^{n}+2\right),\\
N_{n}^{\text{2D}} & =\frac{3}{2}\left(3^{n}-1\right),
\end{split}
\label{eq:-4}
\end{gather}
where we have $n\geq1$ and $n\geq0$ for the 1D and 2D case respectively~\footnote{1D scheme has no meaningful tripartite entanglement at the elementary level since it contains only a bipartite link, therefore $n\geq1$ (see also Fig.~\ref{Fig. 1_SM}[d]).}. The fidelity associated with the network state at the $n^{\text{th}}$ repeater level $\rho_{n}^{\text{1D(2D)}}$
is given by
\[
F_{n}^{\text{1D(2D)}}=\sqrt{\left\langle \Psi_{\rm GHZ}\right\vert \rho_{n}^{\text{1D(2D)}}\left\vert \Psi_{\rm GHZ}\right\rangle },
\]
with $\left\vert\Psi_{\rm GHZ}\right\rangle$ defined in Eq.~\eqref{Eq.GHZ}. The main sources of errors reducing the efficiency of entanglement swapping operations are (i) imperfect memory read-out  (ii) imperfect photodetectors efficiency and (iii) dark counts. These imperfections are associated with an excitation loss probability $v$ during the read-out process, an excitation loss probability $f$ during the photon detection process, and a dark count probability $d$, which is given by the product of the of the dark count rate and the duration of the photon pulses.
The model used for simulating imperfect entanglement swapping operations is explained in detail in Sec.~\ref{sec:Methods}.

To examine the accumulation of errors caused by imperfect swapping operations we consider a drop of fidelity of the entangled states generated by the networks from ideal states of elementary segments (triangles and links for the 2D and 1D schemes, respectively) as a function of total number of swapping operations in the limit of perfect quantum memory. The effect of memory imperfections is discussed in detail below.
\begin{figure}
\includegraphics{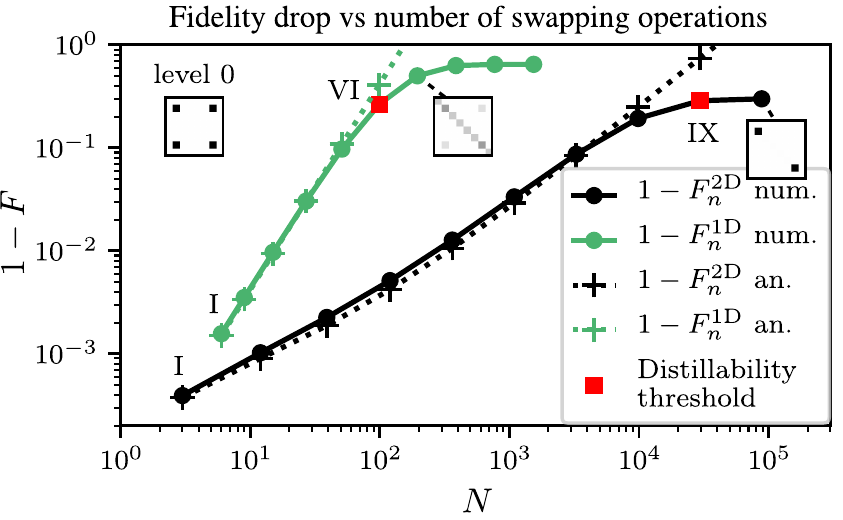}
\caption{Drop of fidelity of entangled states generated by 1D benchmark (green lines) and 2D (black lines) repeater schemes versus total number of imperfect swapping operations
$N_{n}$ [see Eqs.~\eqref{eq:-4}].
Bullets represent numerical data and crosses show analytical results obtained using Eqs.~\eqref{eq:-5} and
\eqref{eq:-6}. Red squares represent the highest nesting
level of the networks generating distillable states. Roman numerals indicate the
number of nesting levels. The insets illustrate density matrices of the generated states
projected onto the qubit basis; the inset labelled "level 0'' represents the
ideal GHZ state, and the insets linked to the numerical points represent
the networks states at the corresponding nesting levels. The parameters for the simulations are
 $v=0.025$, $f=0.025$ and $d=10\text{Hz}\cdot10^{-4}\text{s}=0.001$.}
\label{Fig. 4} 
\end{figure}
The results, shown in Fig.~\ref{Fig. 4}, clearly indicate that the built-in filtering mechanism of the 2D repeater scheme leads to a significantly slower growth of the infidelity with increasing number of swapping operations than the 1D benchmark strategy. These results have been obtained numerically by truncating the considered Hilbert space to include up to four excitations, i.e. up to the Fock state $\text{\ensuremath{\left\vert 4\right\rangle }}$. 
To illustrate the operation of the filtering mechanisms for the analyzed 1D and 2D strategies, we show the initial density matrix (corresponding to an ideal GHZ state)
in the left inset of Fig.~\ref{Fig. 4}, and the density matrices for final states for the 1D (green line inset) and the 2D (black line inset) approach. All density matrices in the insets are projected onto the qubit subspace, which is a good approximation for representation of the final network states. For the 2D repeater scheme, we mainly observe a decay of the
coherences (dephasing), while the 1D benchmark strategy leads not only to dephasing, but also adds significant population to the diagonal elements of the density matrix. Therefore the fidelity is higher in the 2D case for each nesting level, even though the number of required entanglement swapping operations increases faster than in the 1D benchmark approach. Figure~\ref{Fig. 4} also demonstrates that the 2D scheme allows one to implement more repeater nesting levels before the generated  entangled states become not distillable (see Sec.~\ref{sec:Distillability-of-a} for details on the distillation criterion). 

The effect of imperfect swapping operations on the entanglement distribution can be described analytically in the limit in which the corresponding probabilities are small, $v,f,d\ll1$. In this case, the entangled states that are distributed by the network assuming ideal elementary input states, are faithfully represented in the qubit basis ($\ket{0}$ and $\ket{1}$). However, the excited state $\ket{2}$ still needs to be considered perturbatively to account for the scattering of two photons (each from a different memory cell) into one output port of the beamsplitter, which can lead to a single click due to a finite detector efficiency.

Under these assumptions, the resulting states generated by the repeater networks can be presented in terms of the target GHZ state $\rho_{\rm GHZ}=\ket{\Psi_{\rm GHZ}}\bra{\Psi_{\rm GHZ}}$, the corresponding classically correlated state
\[
\rho_{\rm d}=\frac{1}{2}\left(\left\vert 0_{A}0_{B}1_{C}\right\rangle \bra{0_{A}0_{B}1_{C}}+\left\vert 1_{A}1_{B}0_{C}\right\rangle \bra{1_{A}1_{B}0_{C}}\right),
\]
and the diagonal density matrix $\rho_{\mathbb{I}}=\frac{1}{8}\mathbb{I}_{2}\otimes\mathbb{I}_{2}\otimes\mathbb{I}_{2}$,
where $\mathbb{I}_{2}$ is the identity matrix in the qubit basis.

The state generated by the 1D benchmark strategy at the $n^{\rm th}$ nesting level is gviven by
\begin{gather}
\begin{split}\rho_{n}^{\text{1D}} & \approx\left(1-\alpha_{n}\right)\rho_{\rm GHZ}+\alpha_{n}\left(\frac89\rho_{\mathbb{I}}+\frac19\rho_{\rm d}\right),\\
\alpha_{n} & =2\,(N_{n}^{\text{1D}})^{2}\cdot(f+v)d.
\end{split}
\label{eq:-16}
\end{gather}
The 2D repeater scheme generates the following state: 
\begin{gather}
\begin{split}\rho_{n}^{\text{2D}} & \approx\left(1-\alpha_{n}^{\mathbb{I}}-\alpha_{n}^{\rm d}\right)\rho_{\rm GHZ}+\alpha_{n}^{\mathbb{I}}\rho_{\mathbb{I}}+\alpha_{n}^{\rm d}\rho_{\rm d},\\
\alpha_{n}^{\mathbb{I}} & =16\,n\cdot(f+v)d,\\
\alpha_{n}^{\rm d} & =(2N_{n}^{\text{2D}}-4n)\cdot(f+v)d.
\end{split}
\label{eq:-17}
\end{gather}
The infidelities corresponding to the states given by Eqs.~\eqref{eq:-16} and
\eqref{eq:-17} read
\begin{align}
1-F_{n}^{\text{1D}} & =\frac{5}{6}(N_{n}^{\text{1D}})^{2}\cdot(f+v)d\nonumber \\
 & \sim\frac{15}{8}4^{n}\cdot(f+v)d,\label{eq:-5}\\
1-F_{n}^{\text{2D}} & =\frac{1}{2}(N_{n}^{\text{2D}}+12\,n)\cdot(f+v)d\notag\\
 & \sim\frac{3}{4}3^{n}\cdot(f+v)d,\label{eq:-6}
\end{align}
where we used Eqs.~\eqref{eq:-4} to obtain the dominant
scaling of the infidelities with the nesting level $n$. The analytical results [Eqs.~\eqref{eq:-5} and \eqref{eq:-6}] are in a good agreement with the numerical data as shown in Fig.~\ref{Fig. 4} by the green (1D benchmark scheme) and black (2D scheme) dashed lines. The small deviation from the numerics for the 2D repeater curve is due to terms of higher order in the error probabilities~$f$~and~$v$.

The error filtering mechanism is evident from the analytical expressions [Eqs.~\eqref{eq:-16} and \eqref{eq:-17}] for the states generated by both schemes. The lack of contributions to first order in the error probabilities indicates the perfect filtering of single errors. In order to pass undetected, the errors have to conspire during the swapping procedure as represented by the terms $\sim vd$ and $\sim fd$ describing the excitation
loss during memory read-out and the photon loss in the detector, respectively, followed
by a dark count during the detection. The amount of these errors grows linearly with the number
of swapping operations for the 2D repeater and quadratically for
the 1D benchmark scheme.

The structure of the generated states is also different for the two network protocols. Dephasing leading to the classically correlated state~$\rho_{\rm d}$ and processes resembling thermal noise resulting in the diagonal state~$\rho_{\mathbb{I}}$ are contributing similarly in the 1D benchmark strategy of entanglement generation.  In contrast, at high nesting levels~$n$, the 2D repeater scheme suffers mainly from the loss of coherence (dephasing) while the process leading to~$\rho_{\mathbb{I}}$ is significantly suppressed by the built-in error filtering. The resilience of dephasing errors originates from the fact that swapping
operations applied to the ideal state~$\rho_{\rm GHZ}$ and to the dephased one~$\rho_{{\rm d}}$ yield the same outcomes, which renders such errors undetectable.

The numerical and analytical results of this section show that the proposed 2D repeater, in contrast to
the 1D benchmark approach, can efficiently filter out errors introduced by the
imperfect swapping operations and, particularly, by photon losses
in the memory read-out operations. As shown in Sec.~\ref{sec:Memory-decoherence},
the decoherence in the memory cells leads to the same kind of excitation loss
as the read-out inefficiency, and thus is also efficiently filtered
out in the 2D scheme. This explains the main result of the paper:
in the presence of realistic imperfections, the proposed 2D scheme
scales significantly better with the network size than its analogs based on
the 1D repeaters.

\section{Generation of elementary GHZ states\label{sec:Elementary-triangle-generation}}

In this section we present a scheme for the probabilistic generation of the
initial entangled states that constitute the elementary segments of the proposed 2D repeater protocol. We
analyze the segment state and its generation time. It is shown that
the use of a nonlinear node B (see Fig.~\ref{Fig. 5}) is beneficial for the creation of initial states of 2D networks with high rate and fidelity.

\subsection{Scheme\label{subsec:GHZ_generation_Scheme}}

\begin{figure}
\includegraphics{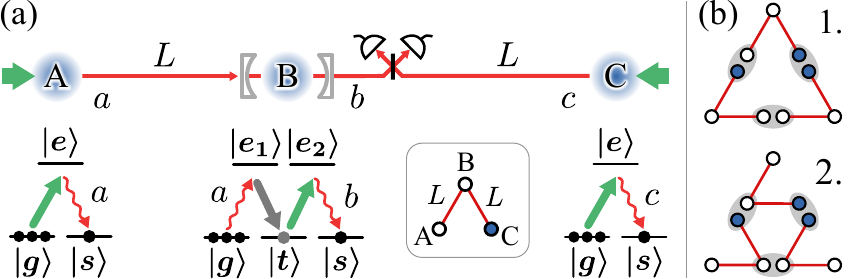}

\caption{(a) Scheme for generating elementary tripartite
GHZ states. Three blue circles A, B, and C represent atomic or solid state
ensembles with level configurations depicted underneath. The corresponding
out-coming photon modes $a$, $b$ and $c$ are represented by red lines.
The thick arrows denote driving laser fields. Node B contains an ensemble with a double-$\Lambda$ scheme in a cavity, which preforms the nonlinear gate operation (see text) in two successive steps illustrated by gray and green arrows in the corresponding level scheme. The swapping station consists of a balanced
beamsplitter and two single photon detectors. The inset schematically
shows the target GHZ state, where the red lines represent the physical
channels; the qubits are of the same (different) colors if they are correlated (anti-correlated). (b) Examples of possible relative orientations of the elementary segments within the $1^{\rm st}$ nesting level of a 2D network;  merging operations are shown by gray ellipses [see Fig.~\ref{Fig. 1_SM}(a)].}
\label{Fig. 5} 
\end{figure}

Our protocol for generating the elementary GHZ states is
inspired by the original DLCZ proposal. The elementary network segment, shown
in Fig.~\ref{Fig. 5}(a), consists of three nodes A, B and C. For simplicity we assume an equilateral triangle with length $L$.
The nodes A and C contain atomic or solid state ensembles with a $\Lambda$-type
level structure, and node B, considered in detail in Sec.~\ref{sec:Double--scheme-ensemble},
employs a cold ensemble with a double-$\Lambda$ configuration placed
in a cavity. Other resources for the elementary state
generation include lasers, fibers, a balanced beamsplitter and two
single photon detectors (SPDs).

The ensembles store quantum information encoded in the absence or presence
of a collective excitation, i.e. in the logical states $\smash{|0\rangle=|g_{1}g_{2}\ldots g_{N}\rangle}$
and $|1\rangle=\sum_{i=1}^{N}|g_{1}g_{2}\ldots s_{i}\ldots g_{N}\rangle/\sqrt{N}$,
where $N$ is the number of emitters per ensemble and $\ket{g}$,
$\ket{s}$ denote the emitter ground states, as shown in Fig.~\ref{Fig. 5}(a). The ensembles can be used to efficiently generate and retrieve excitations, as shown in~\cite{Duan2001,Hammerer2010}.

First, we consider an idealized case of the GHZ state generation to illustrate the working principle of the protocol. The effect of realistic imperfections will be discussed in the next subsection. The target state for the elementary
segment preparation reads $\left\vert \Psi_{\rm GHZ}\right\rangle =(\left\vert 1_{A}1_{B}0_{C}\right\rangle +\left\vert 0_{A}0_{B}1_{C}\right\rangle )/\sqrt{2}$,
where subscripts refer to the corresponding ensembles. The protocol is probabilistic and works as follows. For each generation attempt the ensembles are initialized in the logical state $\ket0$. Then, weak laser pulses drive
the ensembles at nodes A and C, coupling them to the outgoing photonic
modes $a$ and $c$, correspondingly, via an off-resonant Raman transition.
Similarly to~\cite{Duan2001}, the resulting states are two-mode squeezed
vacuum states 
\begin{align}
\left\vert \Psi_{Aa\left(Cc\right)}\right\rangle  & =\text{sech}\,r_{a(c)}\sum_{n=0}^{\infty}\left(\text{tanh}\,r_{a(c)}\right)^{n}\left\vert n_{A\left(C\right)}\right\rangle \left\vert n_{a\left(c\right)}\right\rangle \nonumber \\
 & \propto\sum_{n=0}^{\infty}\epsilon_{a(c)}^{n}\left\vert n_{A\left(C\right)}\right\rangle \left\vert n_{a\left(c\right)}\right\rangle ,\label{eq:-13}
\end{align}
where $r_{a(c)}$ is the squeezing parameter controlled
by the driving laser and sech $x=1/$cosh $x$. To simplify the notation, we introduce the abbreviation $\epsilon_{a(c)}\equiv\text{tanh}\,r_{a(c)}\in[0,1)$.
In the ideal case we choose $\smash{\epsilon_{a}=\epsilon_{c}\equiv\epsilon}$.

Next, both of the outgoing photon pulses propagate over the distance $L$ towards node B, such
that mode $c$ is directed to the central swapping station, and
mode $a$ enters node B. As illustrated in Fig.~\ref{Fig. 5}(a),
node B contains an ensemble with a double-$\Lambda$ configuration, which allows to implement a nonlinear gate operation $\ket{1_{a}0_{B}0_{b}}\rightarrow\ket{0_{a}1_{B}1_{b}}$,
$\ket{0_{a}0_{B}0_{b}}\rightarrow\ket{0_{a}0_{B}0_{b}}$ for the single
photon component of the incoming mode $a$ (see Sec.~\ref{sec:Double--scheme-ensemble}).
The gate action transfers the whole system to a state $\ket{ \Psi_{ABb}}\otimes\ket{\Psi_{Cc}}$, where
\begin{equation}
\left\vert \Psi_{ABb}\right\rangle \propto\left\vert 0_{A}0_{B}\right\rangle \left\vert 0_{b}\right\rangle +\epsilon\left\vert 1_{A}1_{B}\right\rangle \left\vert 1_{b}\right\rangle +\mathcal{O}\left(\epsilon^{2}\right)\label{eq:-7}
\end{equation}
and 
\begin{equation}
\left\vert \Psi_{Cc}\right\rangle \sim\left\vert 0_{C}\right\rangle \left\vert 0_{c}\right\rangle +\epsilon\left\vert 1_{C}\right\rangle \left\vert 1_{c}\right\rangle +\mathcal{O}\left(\epsilon^{2}\right).\label{eq:-7-1}
\end{equation}

Finally, mode $b$ interferes with mode $c$ at the balanced
beamsplitter with the output fields measured by two SPDs.
Upon the detection of a single photon, the elementary segment state is projected onto the
GHZ state $\rho_{\rm e}^{\text{ideal}}=\left\vert \Psi_{\rm GHZ}\right\rangle \left\langle \Psi_{\rm GHZ}\right\vert $, otherwise the generation step is repeated.
In the ideal case, the success probability of a generation attempt
is $q_{1}^{\text{ideal}}=2({\rm sech}\,r~{\rm tanh}\,r)^2\leq1/2$.

A finite efficiency $\eta$ of the nonlinear gate results in the optimal relation between the squeezing parameters $\epsilon_{c}=\sqrt{\eta}\epsilon_{a}$ which maximizes the fidelity of the generated state~$\rho_{\rm e}$. This
relation ensures that there are equal probabilities of detecting a photon coming
from the nodes A and C provided that the chances of the photon loss during the propagation are also equal. For higher nesting levels there are more factors affecting the final state such that a numerical optimization of the squeezing parameters is necessary to achieve the maximum fidelity shown in
Fig.~1 of the main text.

The elementary segment scheme for the 1D DLCZ repeater (Fig.~\ref{Fig. 1_SM}) can be obtained from the 2D scheme shown in Fig.~\ref{Fig. 5}(a) by removing node B and by choosing $\epsilon_{a}=\epsilon_{c}$. In this case the ideal state generated by the scheme is $(\left\vert 1_{A}0_{C}\right\rangle +\left\vert 0_{A}1_{C}\right\rangle )/\sqrt{2}$.

\subsection{Elementary state generation with realistic imperfections\label{subsec:Analysis}}

In this subsection we study the state $\rho_{\rm e}$
of the elementary segment generated in the presence of realistic imperfections.
The most important imperfections in the considered scenario are photon losses in the fiber with attenuation length $L_{\text{att}}$,
photon losses in detectors with probability$f$, dark counts with 
probability $d$, and the finite efficiency $\eta$ of the nonlinear gate at node B (see Sec.~\ref{sec:Double--scheme-ensemble}). Since the memory decay does not significantly affect the elementary segment preparation in the considered parameters regime ($L\lesssim 10^2\,{\rm km}$ and $T_{\rm coh}\gtrsim10^{-2}\,{\rm s}$), we neglect it in this subsection. This assumption allows us to use the simple optimal relation for the squeezing parameters $\epsilon_{c}=\sqrt{\eta}\epsilon_{a}$ maximizing the fidelity of the elementary segment. In the numerical results of the main text, however, we fully account for the state decay in the memory during the propagation time $L/v_{c}$, the detection time given by the pulse duration $\tau$, and the propagation time of classical communication $L/v_{c}$, where $v_c$ is the speed of light in the fiber.

Below we provide an analytical calculation of the state generated by the protocol presented in the previous subsection with the realistic imperfections modeled according to Sec.~\ref{sec:Methods}. The probability of a dark
count $d$, and the probability of a loss in a detector $f$, are treated as small
parameters and are taken into account up to first order. The squeezing parameter $\epsilon_a$ [see Eq.~(\ref{eq:-13})] is also considered to be small and considered up to $\mathcal{O}(\epsilon_a^4)$. Under these assumptions the density matrix of the elementary segment reads 
\begin{multline}
\rho_{\rm e}=\frac{1}{q_{1}}\big(2d\left\vert \text{vac}\right\rangle \left\langle \text{vac}\right\vert +q_{\rm GHZ}\left\vert \Psi_{\rm GHZ}\right\rangle \left\langle \Psi_{\rm GHZ}\right\vert \\
+q_{\text{L}}\left\vert \Psi_{\text{L}}\right\rangle \left\langle \Psi_{\text{L}}\right\vert +q_{\text{R}}\left\vert \Psi_{\text{R}}\right\rangle \left\langle \Psi_{\text{R}}\right\vert \big),\label{eq:-8}
\end{multline}

\noindent where the states are given by
\begin{align*}
\left\vert \text{vac}\right\rangle  & =\left\vert 0_{A}0_{B}0_{C}\right\rangle ,\\
\left\vert \Psi_{\rm GHZ}\right\rangle  & =\big(\left\vert 1_{A}1_{B}0_{C}\right\rangle +\left\vert 0_{A}0_{B}1_{C}\right\rangle \big)/\sqrt{2},\\
\left\vert \Psi_{\text{L}}\right\rangle  & =\big(\sqrt{2}\left\vert 2_{A}1_{B}0_{C}\right\rangle +\left\vert 1_{A}0_{B}1_{C}\right\rangle \big)/\sqrt{3},\\
\left\vert \Psi_{\text{R}}\right\rangle  & =\big(\left\vert 1_{A}1_{B}1_{C}\right\rangle +\sqrt{2}\left\vert 0_{A}0_{B}2_{C}\right\rangle \big)/\sqrt{3}.
\end{align*}
Here the vacuum state $\left\vert \text{vac}\right\rangle $ is generated
by a dark count, the state $\left\vert \Psi_{\text{L}}\right\rangle $
is a result of a photon loss in the fiber AB or in the node B (to the left of the swapping station), and the state $\left\vert \Psi_{\text{R}}\right\rangle $
is produced by a photon loss between the node C and
the swapping station. The states $\left\vert \Psi_{\text{L}}\right\rangle $
and $\left\vert \Psi_{\text{R}}\right\rangle $ contain on average
one excitation more than the target GHZ state and thus are results
of so-called multi-excitation errors.

The coefficient $q_{1}$ in Eq.~\eqref{eq:-8} is the probability to succeed
in a generation attempt. The coefficients in front of the
states in the brackets read 
\begin{gather}
\begin{split}q_{\rm GHZ} & =2\eta e^{-\frac{L}{L_{\text{att}}}}[1-f-\left(1+\eta\right)\epsilon_{a}^{2}]\epsilon_{a}^{2},\\
q_{\text{L}} & =3\eta e^{-\frac{L}{L_{\text{att}}}}(1-\eta e^{-\frac{L}{L_{\text{att}}}})\epsilon_{a}^{4},\\
q_{\text{R}} & =3\eta^{2}e^{-\frac{L}{L_{\text{att}}}}(1-e^{-\frac{L}{L_{\text{att}}}})\epsilon_{a}^{4},\\
q_{1} & =2d+q_{\rm GHZ}+q_{\text{L}}+q_{\text{R}}.
\end{split}
\label{eq:-1}
\end{gather}
One can infer from Eqs.~\eqref{eq:-1} (assuming $\smash{d\ll\epsilon_{a}^{2}}$) that the conditional
probability of the multi-excitation errors grows as $\smash{(q_{\rm L}+q_{\rm R})/q_1\sim\mathcal{O}(\epsilon_{a}^{2})}$ and thus it could be suppressed by decreasing $\epsilon_{a}$. However, once $\epsilon_{a}\to0$, the generated state is deteriorated by the vacuum component as the coefficient $2d$ starts to dominate. Physically this corresponds to a laser driving the ensembles so weakly that most of the detectors clicks are caused by dark counts while the ensembles stay in the vacuum state~$\ket{\rm vac}$.
%However, once $\epsilon_{a}\to0$, the assumption breaks down and the vacuum coefficient $2d$ becomes dominant, as the laser driving the ensembles becomes extremely weak such that most of the detectors clicks are caused by dark counts.
Therefore, there is an optimal value for the parameter $\epsilon_{a}$ corresponding to a compromise between the two sources of errors.

\begin{figure}
\includegraphics{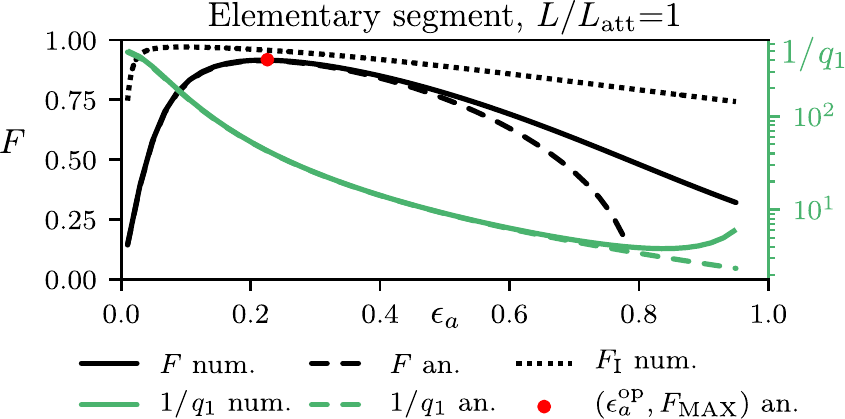}

\caption{Fidelity $F$ of the elementary segment state (black lines), and the
average number of attempts for its generation, $1/q_{1}$ (green
lines), as functions of the parameter $\epsilon_{a}\in[0,1)$ of the
two-mode squeezed state \eqref{eq:-13}. The numerical (num.) data
is calculated in a Hilbert space truncated up to the Fock state $\text{\ensuremath{\left\vert 7\right\rangle }}$,
and the analytical (an.) data are given by Eqs.~\eqref{eq:-1}. The
red dot $(\epsilon_{a}^{\text{op}},F_{\text{MAX}})$ is found analytically using
Eqs.~\eqref{eq:-9}. The dotted line $F_{\text{I}}$ is the numerically
calculated fidelity of the next nesting level, composed by connecting
the corresponded elementary segments in configuration ``$1$.''
of Fig.~\ref{Fig. 5}(b). The representative parameter set is $f=0.05$,
$d=0.001$, and the memory read-out inefficiency $v=0.05$, used for
the nesting level I generation (see Sec.~\ref{sec:Error-filtering-mechanism}).
Memory time $T_{\text{coh}}\to\infty$, and the elementary segment
size is $L/L_{\text{att}}=1$.}
\label{Fig. 6} 
\end{figure}

In Fig.~\ref{Fig. 6} we show the fidelity of the generated state $F =\sqrt{{q_{\rm GHZ}}/{q_{1}}}$ and the average number of the generation attempts $1/q_{1}$ as functions of the squeezing parameter $\epsilon_{a}$. The analytical results are shown as dashed lines. To verify the analytical calculations we present numerical results (solid lines) obtained by the method identical to the analytical analysis but for the truncated Hilbert space including Fock states up to $\ket7$ and without neglecting terms of  higher order in the probabilities $f$ and $d$. This is necessary to study the region of not too small squeezing parameters $\epsilon_a\lesssim1$. As can be seen, the analytical data agree with numerics in the relevant regime of moderate $\epsilon_a$ which provides the maximal fidelity.
The red dot indicates the optimal value $\epsilon_{a}^{\rm op}$ with the corresponding maximal fidelity $F_{\rm MAX}$, which can be found analytically from Eqs.~\eqref{eq:-8} and \eqref{eq:-1}: 
\begin{gather}
\begin{split}\epsilon_{a}^{\text{op}} & \simeq\left(\frac{d\,e^{L/L_{\text{att}}}}{3\eta\left[(1+\eta)/2-\eta e^{-L/L_{\text{att}}}\right]}\right)^{1/4},\\
F_{\text{MAX}} & \simeq\left[1+\frac{\sqrt{6d}}{1-f}\sqrt{e^{L/L_{\text{att}}}(1/\eta+1)-2}\right]^{-1/2},
\end{split}
\label{eq:-9}
\end{gather}
where we also neglected higher orders of $d$.

%To obtain the optimal parameter $\epsilon_{a}$ one can consider dependence
%of the average generation time $T$ and the fidelity $F$ of the elementary
%segment on this parameter, where 
%\begin{gather}
%\begin{split}T & =\Delta t\cdot1/p_{1},\\
%F & =\sqrt{\frac{p_{\rm GHZ}}{p_{1}}}.
%\end{split}
%\label{eq:}
%\end{gather}
%Here $1/p_{1}$ is the average number of the generation attempts and
%$\Delta t=\left(2L+l\right)/v_{c}$ is the time required for one attempt,
%where $L$ is size of the elementary segment, as shown in Fig.~\ref{Fig. 5}(a);
%$l$ is length of the photon pulse, and $v_{c}$ is the speed of light
%(see previous subsection for details).

The state of elementary segments [given by Eq.~\eqref{eq:-8}] that is generated by the scheme with the
nonlinear gate (node B) contains three types of errors. During the merging of the elementary segments into a segment of the $1^{\rm st}$ nesting level these errors can be partially filtered depending on relative orientations of the
elementary segments [a couple of configurations are shown in Fig.~\ref{Fig. 5}(b)]. A proper orientation of segments
leads to the optimal filtering of those errors. Thus, the fidelity
of the $1^{\rm st}$ nesting level state ($F_{\text{I}}$) is generally higher
than the fidelity of the corresponding elementary segments,
as shown in Fig.~\ref{Fig. 6} by the dotted line. The relative orientation of the
elementary segments is optimized numerically for each set of parameters
in order to obtain the fidelity plot shown in Fig.~1 of the main
text.

As can be seen from Fig.~\ref{Fig. 6}, the maximum fidelity $F_{\text{I}}$
of the $1^{\rm st}$ nesting level state is achieved for a
parameter $\epsilon_{a}$ that is smaller than $\epsilon_{a}^{\text{op}}$ corresponding
to the maximum of elementary segment fidelity. This indicates
that the vacuum component of the elementary segments states is more
efficiently filtered out than the multi-excitation error. As presented
in Fig.~5 of the main text, one can achieve an efficient rate/fidelity
trade-off in the limit of long memory coherence times by varying the
parameter $\epsilon_{a}$. However, in the case of a short coherence
time, as shown in the same plot, the time filtering protocol {[}V.
Kuzmin et al., in preparation{]} allows for a more efficient trade-off.

\section{Network generation time\label{sec:Distribution-Time}}

\begin{figure}
\includegraphics{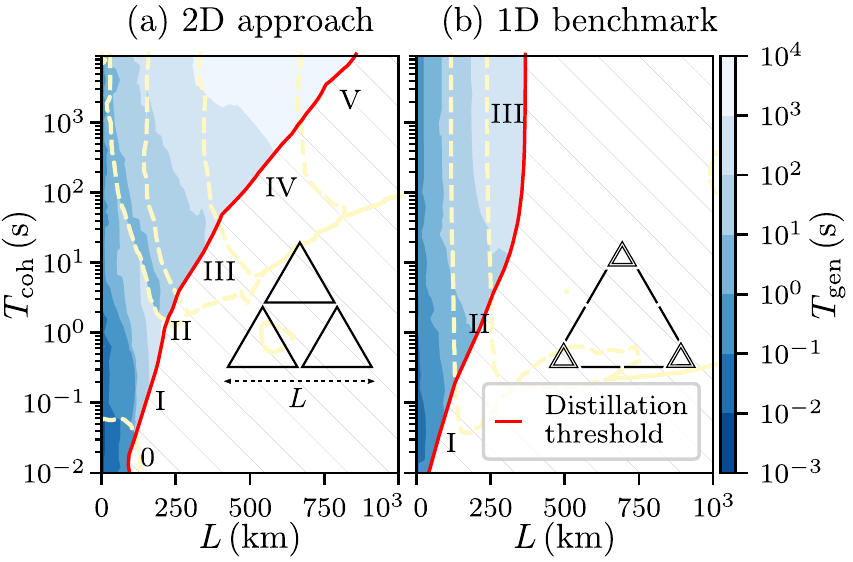}

\caption{Average time $T_{\text{gen}}$ for generating the states with the
maximum achievable fidelity shown in Fig.~1 of the main text:
(a) 2D approach, (b) 1D benchmark approach. Red dash-dotted lines indicate the threshold
where the entanglement distillation is still possible (see Sec.~\ref{sec:Distillability-of-a}).
Dashed lines separate areas with different optimal nesting levels
indicated by Roman numerals. The parameter set includes: the detectors
inefficiency $f=0.05$, dark count probability $d=0.001$, the memories
read-out inefficiency $v=0.05$, and the fiber attenuation length
$L_{\text{att}}=22$ km.}
\label{Fig. 7} 
\end{figure}

%We define the time it takes for the elementary segment generation as $\Delta t\cdot1/q_{1}$, where $q_1$ is the probability to succeed in one generation attempt, such that $1/q_{1}$ is the average number of the generation attempts, and $\Delta t=\tau + 2L/v_{c}$ is the time required for one attempt with $L$ the length of the elementary segment, as shown in Fig.~\ref{Fig. 5}(a); $\tau$ is the photon pulse duration, and $v_{c}$ is the speed of light in the fiber (see previous subsection for details). Calculation of the average time for generation of the total network is explained in Sec.~\ref{sec:Methods}

Figure~\ref{Fig. 7} shows the average time it takes to generate the network state with the maximum fidelity shown in Fig.~1 of the main text. The corresponding calculation is explained in Sec.~\ref{sec:Methods}. One can see that the 2D repeater scheme reaches higher fidelities and greater distribution distances, relative to the 1D benchmark scheme, in exchange for a longer generation time. This is the consequence of the superior filtering mechanism built into the 2D scheme, however, being a passive technique, it still faces a rate-fidelity trade-off.

Multiplexing of the 2D network structure~\cite{Collins2007} allows to significantly reduce the generation time and, simultaneously, to relax the memory coherence time requirements. Meanwhile,
even with multiplexing, the 1D benchmark scheme is unable to achieve the distribution range and fidelity of the 2D repeater network.

\section{Distillability of a GHZ state\label{sec:Distillability-of-a}}

One of the characteristics considered in the main text is the
maximum distance at which the repeater network can distribute quantum states that are
distillable to the perfect GHZ state. This maximum
distance, the distillation threshold, is obtained via the distillation
criterion~\cite{Dur1999} and is shown in Fig.~\ref{Fig. 7}
and Fig.~1 of the main text indicated by red dash-dotted lines.

The idea of the criterion is the following. We consider a family of three-qubit (A, B, and C)  states of the form
\begin{multline}
\rho_{3}=\sum_{\sigma=\pm}\lambda_{0}^{\sigma}\left\vert \psi_{0}^{\sigma}\right\rangle \left\langle \psi_{0}^{\sigma}\right\vert \\
+\sum_{j=1}^{3}\lambda_{j}\left(\left\vert \psi_{j}^{+}\right\rangle \left\langle \psi_{j}^{+}\right\vert +\left\vert \psi_{j}^{-}\right\rangle \left\langle \psi_{j}^{-}\right\vert \right),
\label{eq:-993}
\end{multline}
where the parameters $\lambda_{0}^{\sigma}$ and $\lambda_{j}$ are positive numbers restricted by tr$(\rho_{3})=1$, and
\[
\left\vert \Psi_{j}^{\pm}\right\rangle \equiv\frac{1}{\sqrt{2}}\left(\left\vert j\right\rangle _{AB}\left\vert 1\right\rangle _{C}\pm\left\vert 3-j\right\rangle _{AB}\left\vert 0\right\rangle _{C}\right)
\]
is the orthonormal tripartite GHZ-basis, where $\left\vert j\right\rangle _{AB}=\left\vert j_{1}\right\rangle _{A}\left\vert j_{2}\right\rangle _{B}$
with $j=j_{1}j_{2}$ in binary notation. If the state $\rho_3$ has a negative partial
transposes with respect to the qubits A and B, the maximally entangled
bipartite state, $\left\vert \Phi_{AB}\right\rangle $, can be distilled
from many copies of $\rho_3$. This automatically means that if all
three partial transposes are not positive, one can distill the states
$\left\vert \Phi_{AB}\right\rangle $, $\left\vert \Phi_{AC}\right\rangle $,
$\left\vert \Phi_{BC}\right\rangle $, and generate the target GHZ state
using these states [i.e., as it is shown in Fig.~\ref{Fig. 1_SM}(c)]. Since states can be depolarized to the form \eqref{eq:-993} without changing the diagonal elements~\cite{Dur1999}, this gives a sufficient condition for distillability for general states.

Explicitly, to decide whether the GHZ state $(\left\vert 1_{A}1_{B}0_{C}\right\rangle +\left\vert 0_{A}0_{B}1_{C}\right\rangle )/\sqrt{2}$
is distillable from an ensemble of states $\rho$, one needs to calculate
the values 
\begin{align*}
\lambda_{0}^{\pm} & \equiv\left\langle \Psi_{0}^{\pm}\right\vert \rho\left\vert \Psi_{0}^{\pm}\right\rangle ,\\
2\lambda_{j} & \equiv\left\langle \Psi_{j}^{+}\right\vert \rho\left\vert \Psi_{j}^{+}\right\rangle +\left\langle \Psi_{j}^{-}\right\vert \rho\left\vert \Psi_{j}^{-}\right\rangle.
\end{align*}
The state $\rho$ has a negative
partial transpose $\rho^{T_{A(B,C)}}<0$ with respect to the qubit
A (B, C) iff $|\lambda_{0}^{+}-\lambda_{0}^{-}|-2\lambda_{2(1,3)}>0$.
Thus the sufficient criterion for distillability is
\begin{equation}
|\lambda_{0}^{+}-\lambda_{0}^{-}|-\underset{j=\{1,2,3\}}{\text{max}}\left(2\lambda_{j}\right)>0.\label{eq:-10}
\end{equation}

\section{Methods\label{sec:Methods}}

This section explains two methods we use for simulating the considered
quantum networks. First, we explain the Monte Carlo (MC) method
in the subsection \ref{subsec:Monte-Carlo-method}, and then, in
the subsection \ref{subsec:Diagrammatic-method}, we  provide an overview to the
diagrammatic technique developed in this work.

\subsection{Monte Carlo method\label{subsec:Monte-Carlo-method}}

The idea of the MC method is to simulate numerically the full repeater
protocol step by step. Each run of the program generates a trajectory
with the corresponding total generation time and the final pure state
of the network. Averaging over many runs we obtain the density matrix of the ensemble of states generated by the network and the corresponding mean generation time.

Throughout the simulation, the pure state $\left\vert \Psi\right\rangle $ of the network
evolves in discrete steps describing attempts of the elementary segments preparation and the merging operations. The imperfections of the merging
operations along with the dissipative processes between the merging
steps are described by the corresponding superoperators acting on the system density matrix. A superoperator
can be decomposed into a set of Kraus operators $\left\{ E_{k}\right\}$
which satisfies the completeness relation $\sum_{k}E_{k}=\mathbb{I}$~\cite{Nielsen2010}.
The result of a superoperator action on a pure state $\left\vert \Psi\right\rangle $
(evolution of the state) is an ensemble of states $\{E_{k}\left\vert \Psi\right\rangle \}$
with the corresponding probabilities $p_{k}=\bra{\Psi}E_{k}^{\dagger}E_{k}\ket{\Psi}$.
The MC algorithm randomly chooses a state from the ensemble according
to the probability $p_{k}$.

For the network simulation we take imperfections such
as excitation losses and dark counts into account, which are described by the following
superoperators 
\begin{align}
\mathcal{S}_{\text{loss}}\left(g\right)\bullet & \equiv\text{exp}\left\{ g\,\mathcal{D}\left[a\right]\right\} \bullet,\label{eq:-19}\\
\mathcal{S}_{\text{dc}}\left(d\right)\bullet & \equiv\text{exp}\left\{ d\,\mathcal{D}[a]+d\,\mathcal{D}[a^{\dagger}]\right\} \bullet.\label{eq:-20}
\end{align}
Here $a$ and $a^{\dagger}$ are the operators of annihilation and
creation of an excitation in the corresponding mode, $\mathcal{D}[a]\bullet \equiv a\bullet a^{\dagger}-\frac{1}{2}\left(a^{\dagger}a\bullet+\bullet a^{\dagger}a\right)$
is the Lindblad superoperator. The process of photon losses in the
fiber is described by $\mathcal{S}_{\text{loss}}\left(g\right)$ with
the coefficient $g=L/L_{\text{att}}$, where $L$ is the fiber length
and $L_{\text{att}}$ is the fiber attenuation length; the memory
decay process is defined also by the loss superoperator with the coefficient
$g=T/T_{\text{coh}}$, where $T$ is the decay time and $T_{\text{coh}}$
is the memory coherence time.

Using the introduced imperfections, one can define a superoperator
for the process of merging segments. As shown in Fig.~\ref{Fig. 1_SM}(a),
the merging operation is applied to two adjacent memory modes with
indexes $i$ and $j$. States of the two memories are read out with
the loss probability $v$. Subsequently, the photons interfere in the perfect balanced beamsplitter to
be measured by two photon-number resolving single-photon detectors (SPDs) with the loss
probability $f$ and the dark count probability $d$. The detection of a single photon projects the joint system onto an entangled
state. Altogether, the merging superoperator reads 
\begin{multline}
\mathcal{M}\bullet \equiv 2\left\langle 1_{i}0_{j}\right\vert \{(\mathcal{S}_{\text{det}}^{i}\otimes\mathcal{S}_{{\rm det}}^{j})\mathcal{S}_{\text{BS}}^{ij}(\mathcal{S}_{\text{read}}^{i}\otimes\mathcal{S}_{{\rm read}}^{j})\bullet\}\left\vert 1_{i}0_{j}\right\rangle ,\label{eq:-21}
\end{multline}
where the $i^{\text{th}}$ node memory read out process is described
by $\mathcal{S}_{{\rm read}}^{i}\equiv\mathcal{S}_{\text{loss}}^{i}\left(-\text{ln}[1-v]\right)$.
The dark counts and inefficiency of the SPD at the output port of
the beamsplitter are represented by $\mathcal{S}_{\text{det}}^{i}\equiv\mathcal{S}_{\text{dc}}^{i}\left(d\right)\mathcal{S}_{\text{loss}}^{i}\left(-\text{ln}[1-f]\right)$
with the consecutive projection $2\left\langle 1_{i}0_{j}\right\vert \bullet\left\vert 1_{i}0_{j}\right\rangle $
describing the detection of only one photon. The prefactor $2$ takes the photon detection in the second mode $\left\vert 0_{i}1_{j}\right\rangle $ into account,
which gives the opposite phase of the resulting state and could be
reduced to the first case by the corresponding phase flip operation. The superoperator
for the balanced beamsplitter reads
\begin{align*}
\mathcal{S}_{\text{BS}}^{ij} & \bullet\equiv U_{{\rm BS}}^{ij}\bullet\left(U_{{\rm BS}}^{ij}\right)^{\dagger},\\
U_{{\rm BS}}^{ij} & \equiv\text{exp}\left[\frac{\pi}{4}\left(a_{i}^{\dagger}a_{j}-a_{i}a_{j}^{\dagger}\right)\right],
\end{align*}
where $a_{i(j)}$ is the annihilation operator of the mode $i(j)$.

The numerical calculations are performed in a truncated Hilbert space.
The results presented in the main text account for the Fock states up to $\left\vert 2\right\rangle $, which
is enough to consider the first order term in the multi-excitation
error.

To start the simulations one needs the density matrix of
the elementary segment $\rho_{{\rm e}}$ and its generation probability
$q_{1}$ for one attempt. The calculation of $\rho_{{\rm e}}$ and
$q_{1}$ is done in terms of density matrices and detailed in Sec.~\ref{sec:Elementary-triangle-generation}.
The probability for the elementary segment to be generated in $K$
attempts is $q_{K}=q_{1}\left(1-q_{1}\right)^{K-1}$. The MC calculations
are initialized with each elementary segment having an 'age' of $\Delta t\,K$
with probability $q_{K}$ and being in one of the eigenstates of $\rho_{{\rm e}}$
with the conditional probability given by the corresponding eigenvalue of $\rho_{{\rm e}}$.
Here, the duration of one attempt is $\Delta t=2L/v_{\text{c}}+\tau$
with $v_{\text{c}}$ the speed of light in the fiber and $\tau$ the
photon pulse duration. The first term $2L/v_{\text{c}}$ is the time
of the photon pulse propagation over the distance $L$ between a node
and a swapping station plus the time for the response via classical
communication. For simplicity we consider the network scheme with identical
elementary segments having a certain fixed relative orientation 
[two possible orientations are shown in Fig.~\ref{Fig. 5}(b)]. Thus, the MC simulation starts with all
elementary segments initialized in the same ensemble state $\rho_{{\rm e}}$
and with the generation probability $q_{1}$.

The simulation proceeds as follows. If one segment is created at time $t_{1}$ and an adjacent one at
time $t_{2}>t_{1}$, the nodes of the first one experience decoherence
(decay) for the duration $t_{2}-t_{1}$. Then the two segments are
probabilistically merged by the superoperator $\mathcal{M}$. If the
merging is successful, the state and generation time of the resulting network
segment are saved, and the adjacent segment of the same nesting level
is evaluated. After that, the oldest segment decays for
the difference of the generation times, and the segments are merged
again. Once the swapping fails, the states of the corresponding segments have to be regenerated anew, while the global time keeps running.

The network simulation is performed recursively. To evaluate a segment state
at the $n^{\text{th}}$ nesting level, segments of the level $n-1$
are required, and so on down to the elementary segments. All states
and ages of the intermediate segments are collected for the reconstruction
of their density matrices and generation rates. Thus, as a result of
the MC simulation, the statistics of all nesting levels up to the
$n^{\text{th}}$ one is obtained.

Although the MC method is very flexible, it is extremely demanding
in terms of time and computational resources: one needs to collect
a large number of trajectories in order to achieve convergence to
the true average values.\textcolor{black}{{} In fact, the time for simulation
of one trajectory is proportional to the real network generation time
and grows exponentially with the nesting level $n$. }It could happen
that certain initial parameters lead to an arbitrary long trajectory
to be simulated, and therefore, to the occupation of a large amount of
computational resources. This could be avoided by using a so called
Russian Roulette method~\cite{KahnHerman} which consistently terminates
those trajectories and accounts for them in an unbiased way.

In view of a large computational time, the MC method becomes unpractical for simulation and successive optimization
of scalable and complex networks, such as the 2D repeater network.
We solve this problem by developing a new semi-analytical
diagrammatic method, which significantly simplifies the simulation of
probabilistic quantum repeaters.

\subsection{Diagrammatic method\label{subsec:Diagrammatic-method}}

The idea of the diagrammatic method is to determine the full repeater
network statistics, which can be used then to obtain the average network
state $\rho$ and the generation time $T$. The statistics of states
generated by the network can be described by the density matrix distribution
$\varrho(t)$, such that $\varrho(t)dt$ is the unnormalized density
matrix of the ensemble of states generated within the time period
$[t,t+dt)$ and $\text{Tr}\varrho(t)dt$ is the probability to generate
a state within this period. In what follows we will find the Laplace
transform of the density matrix distribution denoted as
$\tilde{\varrho}(s)$. Then, one can infer the average state generated
by the network and the generation time as

\begin{gather}
\begin{aligned}\rho & =\int_{0}^{\infty}\varrho(t)dt=\tilde{\varrho}(s)|_{s=0},\\
T & =\int_{0}^{\infty}t\,{\rm Tr}\varrho(t)dt=\mathrm{-}{\rm Tr}\left\{ \frac{d\tilde{\varrho}(s)}{ds}\Big|_{s=0}\right\} .
\end{aligned}
\label{eq:-23}
\end{gather}

To illustrate the approach we provide a simple example of the network
evaluation without consideration of the communication time and the
time filtering protocol. We consider a two-segments
repeater shown in Fig.~\ref{Fig. 1_SM}(a), where two links
with states $\rho_{1}$ and $\rho_{2}$ are merged to a longer link
with the state $\rho$. We assume, that the links are generated
probabilistically in discrete time steps $\Delta t$ and that the
success probability in each step is small~$\smash{q_{1}\ll1}$. Under this assumption one can define the link generation rate $\nu=q_{1}/\Delta t$
and the continuous probability density $p_i\left(t\right)=\nu e^{-\nu t}$
to generate the $i^\mathrm{th}$ link at time $t$. Then, we can introduce the elementary
diagram representing the density matrix distribution for generation
of the $i^\mathrm{th}$  elementary link in the state $\rho_i$ at time $t$ 
\begin{equation}
\varrho_{i}(t)=p_i\left(t\right)\rho_{i}=\nu e^{-\nu t}\rho_{i}\equiv\diagram{diag1}.\label{eq:-11}
\end{equation}
Here the vertical line of length $t$ represents the total link generation
time and the circle denotes the successful generation event. Time flows upward.

Using diagram \eqref{eq:-11}, one can represent the density matrix
distribution for two links, $\rho_{1}$ and $\rho_{2}$, prepared
for merging at time $t$ as 
\begin{gather}
\varrho_{\text{prep}}(t)=\diagram{diag2}+\diagram{diag3}\label{eq:-24}\\
=\sum_{\alpha\beta}p_\beta(t)\int_{0}^{t}dt_{0}\,p_\alpha(t_{0})e^{\mathcal{L}_{\alpha}(t-t_{0})}\cdot\rho_1\otimes\rho_2\label{eq:-2}\\
=\nu^2 \sum_{\alpha}e^{-\nu t}\int_{0}^{t}dt_{0}\,e^{-\nu t_0}e^{\mathcal{L}_{\alpha}(t-t_{0})}\cdot\rho_1\otimes\rho_2,\nonumber 
\end{gather}
where we sum over two possible generation orders $\left(\alpha,\beta\right)=\{(1,2),(2,1)\}$.
In Eq.~$\eqref{eq:-2}$ we integrate over all possible preparation
times $t_{0}$ of the $1^{\text{st}}$ generated link, $\alpha$,
before the $2^{\rm nd}$ one, $\beta$, is prepared at time $t$. The degradation
of the link $\alpha$ during the waiting time $t-t_{0}$ due to the
finite memories life time is taken into account using the superoperator
$\mathcal{L}_{\alpha}=T_{\text{coh}}^{-1}\left\{ \mathcal{D}\left[a_{1,\alpha}\right]+\mathcal{D}\left[a_{2,\alpha}\right]\right\} $,
where $a_{1(2),\alpha}$ are the annihilation operators for two memory
nodes of the link $\alpha$ and $\mathcal{D}[a]$ is the Lindblad
superoperator defined in the previous subsection.

The Laplace image of the distribution $\eqref{eq:-2}$ reads $\tilde{\varrho}_{\text{prep}}(s)=\sum_{\alpha}\nu^{2}/[(2\nu+s)(\nu+s-\mathcal{L}_{\alpha})]\cdot\rho_1\otimes\rho_2$.
Therefore, the average density matrix and the average time for preparation
of the two links for merging can be found according to Eqs.~$\eqref{eq:-23}$
as 
\begin{gather}
\begin{split}\rho_{\text{prep}} & =\tilde{\varrho}_{\text{prep}}(s)|_{s=0}=\frac{1}{2}\sum_{\alpha}\frac{\nu}{\nu-\mathcal{L}_{\alpha}}\cdot\rho_{1}\otimes\rho_{2},\\
T_{\text{prep}} & =\mathrm{-}{\rm Tr}\left\{ \frac{d\tilde{\varrho}_{{\rm prep}}(s)}{ds}\Big|_{s=0}\right\} =\frac{3}{2\nu}.
\end{split}
\label{eq:-22}
\end{gather}

Once both links are prepared, they are probabilistically merged into
the link $\rho$. This merging can be represented by the modified
diagrams $\eqref{eq:-24}$ as 
\[
\mathcal{M}\varrho_{\text{prep}}(t)=\diagram{diag4}+\diagram{diag5}\equiv\diagram{diag6},
\]
where the operator of the probabilistic merging $\mathcal{M}$ is
defined in Eq.~$\eqref{eq:-21}$.

The probability densities of the successful and unsuccessful merging
events are ${\rm Tr}\,\mathcal{M}\varrho_{\text{prep}}(t)$ and\textcolor{red}{{}
}${\rm Tr}\,(\mathcal{I}-\mathcal{M})\varrho_{\text{prep}}(t)$, correspondingly,
where \textbf{$\mathcal{I}$} is the unit superoperator. To fully
describe the network, one has to consider all possible trajectories
consisting of different numbers of unsuccessful mergings leading to
the last successful one. The density matrix distribution for the network
state generated at time $t$ after one unsuccessful merging is represented
by the following diagram 
\begin{gather}
\diagram{diag7}\nonumber \\
=\int_{0}^{t}dt_{0}\,\mathcal{M}\varrho_{\text{prep}}(t-t_{0})\cdot{\rm Tr}\,(\mathbb{\mathcal{I}}-\mathcal{M})\varrho_{\text{prep}}(t_{0}),\label{eq:-3}
\end{gather}
where we integrate over all intermediate times $t_{0}$ of the unsuccessful
merging. The convolution in Eq.~\eqref{eq:-3} becomes a product
in the Laplace domain: $\mathcal{M}\tilde{\varrho}_{\text{prep}}(s)\cdot{\rm Tr}\,(\mathcal{I}-\mathcal{M})\tilde{\varrho}_{\text{prep}}(s)$.
The density matrix distribution of the final state generated by the
network is described by the following infinite sum of diagrams 
\[
\varrho(t)=\diagram{diag6}+\diagram{diag7}+\diagram{diag8}+\dots
\]
In the Laplace domain the sum becomes a geometric series which converges
to 
\begin{equation}
\tilde{\varrho}(s)=\frac{\mathcal{M}\tilde{\varrho}_{\text{prep}}(s)}{1-{\rm Tr}\,(\mathcal{I}-\mathcal{M})\tilde{\varrho}_{\text{prep}}(s)}.\label{eq:-14}
\end{equation}

According to Eqs.~$\eqref{eq:-23}$ the average density matrix and
the average generation time for the merging of two links are found
from Eq.~\eqref{eq:-14} as 
\begin{align*}
\rho & =\tilde{\varrho}(s)|_{s=0}=\frac{\mathcal{M}\tilde{\varrho}_{\text{prep}}(0)}{\text{Tr}\,\mathcal{M}\tilde{\varrho}_{\text{prep}}(0)},\\
T & =\mathrm{-}{\rm Tr}\left\{ \frac{d\tilde{\varrho}(s)}{ds}\Big|_{s=0}\right\} =\frac{1}{\text{Tr}\,\mathcal{M}\tilde{\varrho}_{\text{prep}}(0)}\cdot\frac{3}{2\nu}.
\end{align*}

The next nesting level of the network can be evaluated by repeating
the described procedure using the obtained state $\rho$ and rate
$\nu=1/T$ as initial parameters for the new ``elementary'' links.

The application of the diagrammatic method for the evaluation of the proposed 2D repeater architecture,
as well as the calculation of the communication time and the temporal
filtering, shown in Fig.~5 of the main text, will be presented
in detail in {[}V. Kuzmin et al., in preparation{]}.

%%%%%%%%%%%%%%%%%%%%%%%%%%%%%%%%%%%%%%%%%%%%%%%%%%%%%%%%%%%%

\section{Memory decoherence\label{sec:Memory-decoherence}}

As discussed in Section \ref{subsec:GHZ_generation_Scheme}, the quantum
information in ensembles is encoded in collective excitations $|0\rangle=|g_{1}g_{2}\ldots g_{N}\rangle$
and $|1\rangle=\frac{1}{\sqrt{N}}\sum_{i=1}^{N}|g_{1}g_{2}\ldots s_{i}\ldots g_{N}\rangle$.
In this section we study the effect of memory decoherence on the collective
excitations. The memory based on ensembles (both, atomic and solid
state) decays mainly due to individual dephasing of atoms (impurities)
which can be caused by atomic collisions or by fluctuating magnetic fields~\cite{Hammerer2010}.

Here we describe the evolution of the collective excitations in terms
of Heisenberg-Langevin equations~\cite{Barnett2002} and show that the individual dephasing
results simply in a decay of the collective variables
\begin{align}
\dot{X} & =-\frac{\gamma}{2}X+\sqrt{\gamma}f_{X}(t),\label{eq:canonical}\\
\dot{P} & =-\frac{\gamma}{2}P+\sqrt{\gamma}f_{P}(t),\label{eq:canonical2}
\end{align}
where we introduce canonical variables for spin polarized ensembles
$X=J_{y}/\sqrt{\langle J_{z}\rangle}$, $P=J_{x}/\sqrt{\langle J_{z}\rangle}$,
with $J_{x,y,z}$ the projections of collective spin. The Langevin
noise operators $f_{X(P)}(t)$ are given by the correlators 
\begin{align*}
\langle f_{X(P)}(t)f_{X(P)}(t')\rangle & =\frac{NS/2}{\langle J_{z}\rangle}\delta(t-t')\\
 & =\frac{1}{2}\delta(t-t')+\mathcal{O}(N^{-1}),\\
\langle[f_{X}(t)f_{P}(t')]\rangle & =i\,\delta(t-t').
\end{align*}
Here $S$ is a single spin value, and $N$ is the number of spins
(atoms).

To obtain the result above, we consider an ensemble of
spins (atoms) which experience individual random rotations around
a certain direction ($z$-axis) due to perturbations caused by collisions and
fluctuating magnetic fields. The preferred direction can be chosen
by applying an extra magnetic field which creates an energy splitting
of the atomic ground levels larger than the energy scale of the perturbation.
The Hamiltonian describing the dephasing effect reads 
\begin{equation}
H=\sqrt{\gamma_{z}}\sum_{i}f_{z}^{(i)}(t)\sigma_{z}^{(i)},\label{app:Hamiltonian_dephase}
\end{equation}
where $f_{z}^{(i)}(t)$ are Langevin forces with $\langle f_{z}^{(i)}(t)f_{z}^{(j)}(t')\rangle=\delta_{ij}\delta(t-t')$.
The individual spin components obey the commutation relations
$[\sigma_{x},\sigma_{y}]=i\,\sigma_{z}$ (plus cyclic permutations).
We are interested in the effective equations of motion for the collective
spin components $J_{x,y,z}=\sum_{i}\sigma_{x,y,z}^{(i)}$. Using the Heisenberg equations
of motion for the individual spins $\sigma_{x,y,z}^{(i)}$ (given
by the Hamiltonian~\eqref{app:Hamiltonian_dephase}) we obtain the equations
for the collective spin,
\begin{align}
\dot{J}_{x} & =-\sqrt{\gamma_{z}}\sum_{i}f_{z}^{(i)}(t)\sigma_{y}^{(i)},\label{eq:collective_spin_x}\\
\dot{J}_{y} & =\sqrt{\gamma_{z}}\sum_{i}f_{z}^{(i)}(t)\sigma_{x}^{(i)},\label{eq:collective_spin_y}\\
\dot{J}_{z} & =0.\label{eq:collective_spin_z}
\end{align}
Next, we formally solve the equations for the individual spins $\sigma_{x,y,z}^{(i)}$
and substitute the solutions into the equations for the collective
variables~\eqref{eq:collective_spin_x}-\eqref{eq:collective_spin_z}.
Using the $\delta$-correlated property of the Langevin noise operators we
obtain the effective equations of motion for the collective spin components,
\begin{align*}
\dot{J}_{x} & =-\frac{\gamma_{z}}{2}J_{x}-\sqrt{\gamma_{z}}\sum_{i}f_{z}^{(i)}(t)\sigma_{y}^{(i)}(0),\\
\dot{J}_{y} & =-\frac{\gamma_{z}}{2}J_{y}+\sqrt{\gamma_{z}}\sum_{i}f_{z}^{(i)}(t)\sigma_{x}^{(i)}(0),\\
\dot{J}_{z} & =0.
\end{align*}
Here $\sigma_{x,y,z}^{(i)}(0)$ are the initial states of atomic spins.
For an initially polarized ensemble $\langle\sigma_{z}^{(i)}(0)\rangle=S$
and using $\langle\sigma_{x(y)}^{2}(0)\rangle=S/2$ and $\langle J_{z}(0)\rangle=NS$
one arrives at the effective equations of motion for the canonical
collective variables \eqref{eq:canonical},~\eqref{eq:canonical2}.
These equations describe a decay of continuous variables with the
minimal quantum noise and corresponding decay rate $\gamma=\gamma_{z}$.
Our numerical simulations employ this model for the quantum memory
decoherence.

Alternatively, one can consider a similar model with individual spins
randomly rotating around all three axes. In this case the equations
\eqref{eq:canonical},~\eqref{eq:canonical2} still hold for $\gamma/2=\gamma_{x}=\gamma_{y}=\gamma_{z}$
and the noise properties become 
\begin{align*}
\langle f_{X(P)}(t)f_{X(P)}(t')\rangle & =\frac{N(S/2+S^{2})}{\langle J_{z}\rangle}\delta(t-t')\\
 & =\left(\frac{1}{2}+S\right)\delta(t-t')+\mathcal{O}(N^{-1}),\\
\langle[f_{X}(t)f_{P}(t')]\rangle & =i\,\delta(t-t').
\end{align*}

\section{Double-$\Lambda$ scheme ensemble\label{sec:Double--scheme-ensemble}}

In this section we consider the operation of the atomic ensemble
at node B [see Fig.~\ref{Fig. 5}(a)] for generating GHZ states at the elementary level of the 2D repeater scheme. Node B acts as a nonlinear
gate which does not affect the input field in the vacuum state but
converts a single input photon into an output photon and one excitation
in the atomic ensemble: $\ket{0_{a}0_{B}0_{b}}\rightarrow\ket{0_{a}0_{B}0_{b}}$,
$\ket{1_{a}0_{B}0_{b}}\rightarrow\ket{0_{a}1_{B}1_{b}}$, where the
subscript $a$~($b$) refers to the incoming~(outgoing) photonic mode.

As shown in Fig.~\ref{Fig. 5}(a), node B consists of an ensemble of atoms with a level scheme in the
double-$\Lambda$ configuration placed in a cavity. The
desired gate operation is performed in two steps. First, node
A emits light which is entangled with ensemble A and is also mode
matched (by tuning the time dependence of the atoms-field coupling)~\cite{Cirac1997,Ritter2012,Gorshkov2007,Fleischhauer2000,Dilley2012,Vogell2017}
to enter the cavity in node B. In node B the
light acts on the atomic transition $|e_{1}\rangle\langle g|$ in
resonance with a detuned laser field driving the transition $|e_{1}\rangle\langle t|$.
Such a $\Lambda$-configuration converts the state of the incoming
field mode $a$ into the collective coherence between the ground states
$|g\rangle$ and $|t\rangle$ of atoms in the ensemble analogous to
a quantum memory for light~\cite{Gorshkov2007}. The process of
the photon storage is enhanced both by the number of atoms $N$
and by the cavity finesse $\mathcal{F}={2\pi}/(\tau\kappa)$, where
$\tau$ and $\kappa$ are the cavity round trip time and the linewidth,
respectively. As shown in~\cite{Gorshkov2007} the efficiency (probability
to store a photon) of the quantum memory is given by $\eta=C/(1+C)$,
where $C={\mathcal{O}_{d}\mathcal{F}}/(4\pi)$ is the cooperativity
with the resonant optical depth $\mathcal{O}_{d}=\sigma N/S$,
the beam cross section $S$, and the resonant scattering cross section
$\sigma$ (for a two-level atom $\sigma=3\lambda^{2}/2\pi$). Due to the cavity enhancement and the large optical depth of the ensembles assumed here, the efficiency of the storage process approaches unity.

Once the incoming photon is stored in the collective coherence $\sum_{i=1}^{N}|g\rangle_{i}\langle t|$,
one can enable the second $\Lambda$-transition to transfer the excitation
to the collective coherence $B=\sum_{i}^{N}|g\rangle_{i=1}\langle s|$
while creating a photon in the outgoing mode $b$. This is achieved,
similarly to the previous step, by driving the transition $|e_{2}\rangle\langle t|$
with a detuned laser. This way, a photon creation in the mode supported by the cavity, coupled to the transition $|e_{2}\rangle\langle s|$ enables
the atomic excitation transfer to the collective polarization $B$.

Analogous to the first $\Lambda$-transition, the efficiency
of the process is given by $\eta=C_{1}/(1+C_{1})$, where $C_{1}$
is the cooperativity corresponding to the single atom since there
is no coherent enhancement by the number of atoms in the ensemble. Thus a cavity with a good cooperativity is required for the nonlinear gate at node B.

Experiments for single atoms reach cooperativities of, e.g., $C_{1}=328$
\cite{Hood2000}, $C_{1}=45$~\cite{Hamsen2017}. For experiments
with atomic ensembles in cavities, single atom cooperativities $C_{1}\sim300$
\cite{Colombe2007} and $C_{1}=105$~\cite{Gupta2007} have been reached.
For solid state ensembles, the combination of such memories with cavities is under current development~\cite{HungerPers}.
In addition, all-fiber cavities~\cite{Kato2015,Schneeweiss2017} are
promised to achieve $C_{1}\sim1$ for cold atomic ensembles. In the
current work we use the value $C_{1}=1.5$ with corresponding efficiency
$\eta=0.6$ of the node B, corresponding near future realistic
parameters. For completeness, a scheme without cavities is discussed in the
next section.

In order to define the superoperator of node B, we consider a
simplified model, which describes its essential features. In this
model, the finite efficiency $\eta$ of the node B is the result of
spontaneous emissions in the second step of the described gate. A
spontaneous emission leads either to population of the wrong atomic
level or to the generation of a wrong spin wave in the ensemble. In both
cases neither the collective coherence $B$ nor an out-coming photon
in the mode $b$ are produced. Thus the error can be described as a
decay to a vacuum state $\ket{1_{a}0_{B}0_{b}}\to\ket{0_{a}0_{B}0_{b}}$.

Based on the given model, the evolution in the node~B
is defined by the superoperator
\[
\mathcal{S}_{B}\bullet=\text{Tr}_{a}\left\{ U\left[S_{\text{loss}}^{a}(-\text{ln}\,\eta)\,\bullet\right]U^{\dagger}\right\} ,
\]
where $S_{\text{loss}}^{a}$ is the superoperator describing losses in mode $a$ and given by Eq.~\eqref{eq:-19}
and $U=\text{exp}[\frac{\pi}{2}(aB^{\dagger}b^{\dagger}-\text{h.c.})]$
is the evolution operator describing the swapping of the incoming
photon into one outgoing photon and one excitation in the ensemble.
The partial trace means that the incoming mode $a$ is not measured
after the interaction. In principle one can measure the mode $a$
in order to filter out erroneous states caused by the higher Fock
components of the incoming mode. However, this erroneous component
is small, $\sim\mathcal{O}\left(\epsilon_{a}^{2}\right)$, while an
imperfect detector, used for the filtering, by itself can introduce
additional errors.

\section{2D repeater scheme without cavities}\label{sec:Cavityless-scheme-of}

\begin{figure}[t]
\includegraphics[width=0.9\columnwidth]{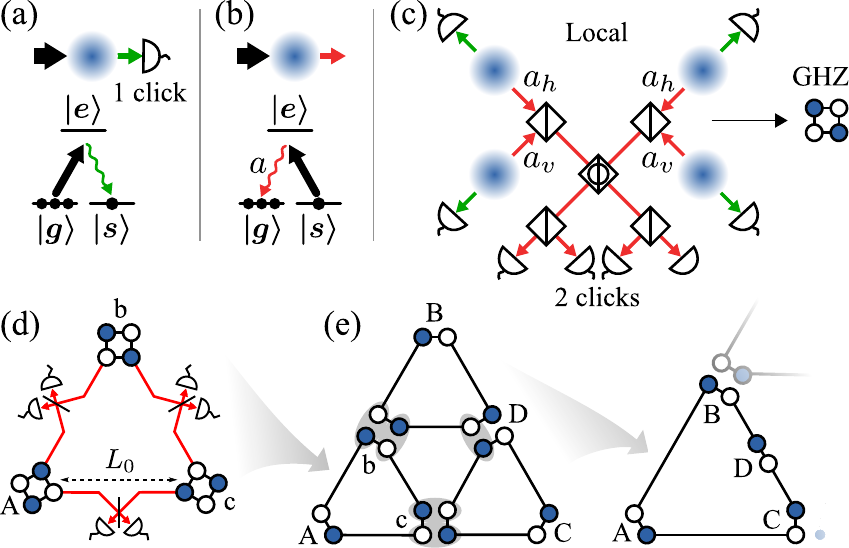}
\caption{2D repeater scheme without cavities. (a) Preparation of a single
photon source: a weak coherent pulse (bold arrow) is scattered off a
three-level ensemble (blue cloud) leading to the emission of a Stokes photons (green arrow).
The memory excitation is created probabilistically conditioned on a single photon detection. (b)
Full or partial read-out of the stored excitation to an anti-Stokes
photon $a$ (red arrow). (c) Scheme for a local four-partite GHZ state
preparation, shown in the picture in notation of Fig.~\ref{Fig. 2_SM}.
The prepared single photon sources partially emit stored excitations
to photonic modes, $a_{h\left(v\right)}$, with horizontal (vertical)
polarization. Squares with vertical bar correspond to polarizing beam
splitters (PBS), and the central square with a circle represents a PBS with a $45^{\circ}$ rotated basis. The successful preparation of the entangled state is conditioned
on detection of two photons on certain detectors as explained in~\cite{Sangouard2008}. (d) Scheme for generating the
elementary segment of length $L_{0}$. (e) Example for entanglement
distribution between four parties: A, B, C, and the intermediate node
D. Gray ellipses represent merging operations [see Fig.~\ref{Fig. 1_SM}(a)].}
\label{Fig. 8} 
\end{figure}
\begin{figure}[b]
\includegraphics{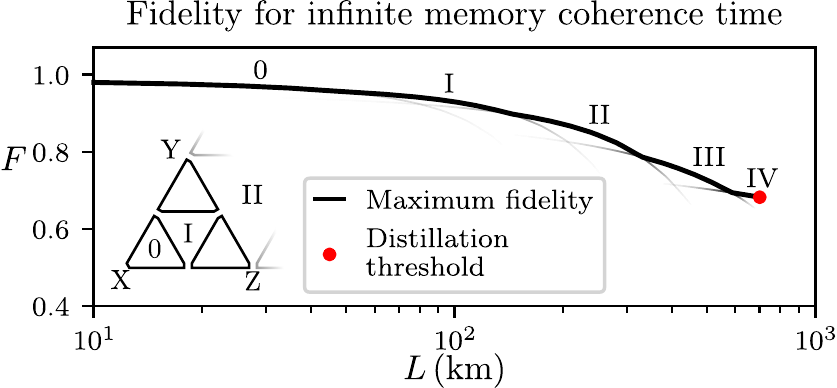}
\caption{Maximum achievable fidelity of a tripartite GHZ state generated by the scheme  without cavities in the infinite memory limit. The fidelity
is found by projecting the resulting network state onto a single excitation subspace, which emulates the final post-selection of
the states with one excitation at each node, analogously to the DLCZ
protocol. Roman numerals indicate the optimal nesting levels. The red dot
represents the point beyond which entanglement distillation is impossible
(see Sec.~\ref{sec:Distillability-of-a}). The imperfection parameters are similar to the ones used in Fig.~4 of the main text: 
detectors inefficiency $f=0.05$, dark counts probability $d=0.001$,
memories read-out inefficiency $v=0.05$, and fiber attenuation
length $L_{\mathrm{att}}=22$~km.}
\label{Fig. 9} 
\end{figure}

In this section we briefly consider an alternative implementation of
our 2D repeater scheme that dispenses with cavities. This scheme, depicted in Fig.~\ref{Fig. 8}, utilizes the same ingredients as
the original DLCZ protocol~\cite{Duan2001}, i.e. linear
optical elements, photon detectors and ensembles. Dispensing with cavities, however, leads to an increase of
the required resources and the generation times, implying that the
scheme demands much longer coherence times for quantum memories. Therefore, implementation of this scheme may be less feasible than the 2D repeater scheme proposed in the main text. However, the alternative scheme has similar fidelity scaling, given in Fig.~\ref{Fig. 9}, what demonstrates robustness of the 2D repeater concept proposed in~\cite{Wallnofer2016}.

The scheme without cavities, shown in Fig.~\ref{Fig. 8}(c), employs a method~\cite{Sangouard2008} for the local generation of high fidelity entangled pairs of atomic
excitations that is based on the preparation of atomic excitations and their partial readout [see Fig.~\ref{Fig. 8}(a,b)].
The prepared entangled pairs are merged by three one-photon swapping
operations [c.f. Fig.~\ref{Fig. 1_SM}(a)] establishing the elementary
long distance entangled states illustrated in Fig.~\ref{Fig. 8}(d).
The entanglement distribution is presented in Fig.~\ref{Fig. 8}(e).

The scheme without cavities exploits polarization-like encoding
where quantum information is stored in states of two local memories. The
logical states 
\begin{align*}
\ket0_{\text{logic}}&\equiv\ket{10}_{\text{mem}},\\
\ket1_{\text{logic}}&\equiv\ket{01}_{\text{mem}},
\end{align*}
are represented by the photon-number encoded memory states 
\begin{align*}
\ket0_{\text{mem}}&\equiv|g_{1}g_{2}\ldots g_{N}\rangle,\\
\ket1_{\text{mem}}&\equiv \frac1{\sqrt{N}}\sum_{i=1}^{N}|g_{1}g_{2}\ldots s_{i}\ldots g_{N}\rangle,
\end{align*}
where $N$ is the number of emitters per ensemble with $\ket{g}$ and $\ket{s}$ the emitter ground states.

Since the logical states are encoded with in two physical memories, the
merging operation can be implemented employing one or two one-photon
swapping operations, as illustrated in Fig.~\ref{Fig. 8}(e). The
former option allows one to include the intermediate node in the distributed
entangled state, shown as node D in the example.

The long generation times, mentioned above, originate from the fact
that the generation of elementary segments of the scheme relies on
several steps of probabilistic photon detections and thus requires
a large number of generation attempts. However, the time for the attempt
could be significantly reduced by exploiting \textcolor{black}{high
bandwidth quantum memories capable of operating with short-length
photon pulses, making operations of photodetection faster.}

To illustrate the performance of the scheme  without cavities, we considered
a tripartite GHZ state distribution in the infinite memory limit.
Fig.~\ref{Fig. 9} presents the maximum achievable fidelity as a
function of distance. The plot demonstrates a high network scalability,
which is very similar to the scalability of the cavity-exploiting
scheme, illustrated in Fig.~4 of the main text. This demonstrates
the efficiency of the proposed ensemble-based implementation scheme
of the 2D repeater networks.

\bibliographystyle{naturemag}
\bibliography{Bibliography}

\end{document}